\PassOptionsToPackage{table,x11names}{xcolor}
\RequirePackage{fix-cm}
\documentclass{svjour3}

\usepackage{natbib}
\let\textcite\citet

\usepackage{xspace}
\usepackage{hyperref}

\usepackage{lmodern}
\usepackage{algorithmic}
\usepackage{graphicx}
\usepackage{textcomp}
\usepackage{colortbl}
\usepackage{tabularx,booktabs}
\usepackage{tcolorbox}
\usepackage{caption,subcaption}
\usepackage{enumerate}
\usepackage[inline]{enumitem}
\usepackage{multirow}
\usepackage{rotating}
\usepackage{pdflscape}
\usepackage{rotating}

\usepackage{listings}

\definecolor{codeblack}{rgb}{0,0,0}
\definecolor{lightgray}{gray}{0.98}
\definecolor{codeblue}{rgb}{0.13,0.13,1}
\definecolor{codegrey}{rgb}{0.36,0.35,0.38}
\definecolor{codegreen}{rgb}{0,0.5,0}
\definecolor{codered}{rgb}{0.9,0,0}
\definecolor{purple}{cmyk}{0.41,0.73,0,0.1}

\makeatletter
\lst@Key{countblanklines}{true}[t]%
    {\lstKV@SetIf{#1}\lst@ifcountblanklines}

\lst@AddToHook{OnEmptyLine}{%
    \lst@ifnumberblanklines\else%
       \lst@ifcountblanklines\else%
         \advance\c@lstnumber-\@ne\relax%
       \fi%
    \fi}
\makeatother

\usepackage[normalem]{ulem}
\definecolor{responseblue}{RGB}{0, 0, 255}
\definecolor{responsered}{RGB}{255,0,0}
\definecolor{responseorange}{RGB}{255,165,0}

\usepackage{apastat}

\newcommand{\highlight}[1]{\cellcolor{Wheat1!40}\textbf{#1}}
\newcommand{\markbox}[2]{\tcbox[size=tight,boxsep=0.5mm,on line,colframe=#1]{#2}}
\newcommand{\two}[1]{\markbox{red}{#1}}

\newlist{rqs}{enumerate}{1}
\setlist[rqs]{label={\textbf{RQ\arabic*}},ref={RQ\arabic*},resume=rqs}

\usepackage{tcolorbox}
\tcbuselibrary{theorems}
\tcbuselibrary{breakable}
\newtcbtheorem[auto counter]{finding}{Finding}{theorem style=plain,fonttitle=\bfseries,breakable=true,label separator={},colframe=black!30,coltitle=black,left=0.5em,right=0.5em,top=0.25em,bottom=0.25em}{}

\makeatletter
\def\cl@chapter{\@elt {theorem}}
\makeatother
\usepackage[nameinlink,capitalize]{cleveref}
\crefname{tcb@cnt@finding}{finding}{findings}
\Crefname{tcb@cnt@finding}{Finding}{Findings}

\usepackage{etoolbox}
\makeatletter
\pretocmd{\NAT@citexnum}{\@ifnum{\NAT@ctype>\z@}{\let\NAT@hyper@\relax}{}}{}{}
\makeatother

\begin{document}

\title{An Exploratory Eye Tracking Study on  How Developers Classify and Debug Python Code in Different Paradigms}
\titlerunning{How Developers Classify and Debug Python Code in Different Paradigms}

\author{Samuel W. Flint\thanks{The majority of this work was completed while S. Flint was at the University of Nebraska--Lincoln.} \and Jigyasa Chauhan \and Niloofar Mansoor \and Bonita Sharif \and Robert Dyer}
\institute{
  Corresponding author: Samuel W. Flint \at
  Dakota State University\\
  \email{Samuel.Flint@dsu.edu}
  \and
  Co-first-author: Jigyasa Chauhan \at
  University of Nebraska--Lincoln\\
  \email{chauhan@huskers.unl.edu}
  \and
  Niloofar Mansoor \at
  University of Nebraska--Lincoln\\
  \email{niloofar@huskers.unl.edu}
  \and
  Bonita Sharif \at
  University of Nebraska--Lincoln\\
  \email{bsharif@unl.edu}
  \and
  Robert Dyer \at
  University of Nebraska--Lincoln\\
  \email{rdyer@unl.edu}
}

\date{Received: date / Accepted: date}

\maketitle

\begin{abstract}
Modern programming languages, such as Python, support language features from several paradigms, such as object-oriented, procedural, and functional.
Research has shown that code written in some paradigms can be harder to comprehend, but to date, no research has looked atwhich paradigm-specific language features impact comprehension.
To this end, this study seeks to uncover which paradigm-specific features impact comprehension and debugging of code or how multi-paradigm code might affect a developer's ability to do so.
We present an exploratory empirical eye-tracking study to investigate 1) how developers classify the predominant paradigm in Python code and 2) how the paradigm affects their ability to debug Python code. The goal is to uncover if specific language features are looked at more often while classifying and debugging code with a predominant paradigm. Twenty-nine developers (primarily students) were recruited for the study and were each given four classification and four debugging tasks in Python. Eye movements were recorded during all the tasks.
The results indicate confusion in labeling Functional and Procedural paradigms, but not Object-Oriented. The code with predominantly functional paradigms also took the longest to complete. Changing the predominant paradigm did not affect the ability to debug the code, though developers did rate themselves with lower confidence for Functional code. We report significant differences in reading patterns during debugging, especially in the Functional code.  During classification, results show that developers do not necessarily read paradigm-relevant token types.
\end{abstract}

\section{Introduction}
\label{sec:intro}

Most modern programming languages have a predominant programming paradigm they support, often object-oriented.  Yet they are also multi-paradigm and support other programming paradigms, such as procedural or functional, in addition to its predominant paradigm.  Python, a popular and currently one of the fastest growing programming languages~\citep{tiobe,pypl,TopProgrammingLanguage}, is a good example of such a multi-paradigm language.  To date, little is known about how developers utilize the different programming paradigms in multi-paradigm languages, such as Python.

Determining which paradigm some code is a part of helps developers select the right data structure(s) and algorithm(s) for the problem at hand. 
Thus, code classification is an important part of how developers comprehend code \citep{Floyd1979}.
Knowing the paradigm used also explains how developers build mental models of the code \citep{BROOKS1983543,LETOVSKY1987325, rist-plans, Soloway80, PENNINGTON1987295,vonmayrhauser95}. 
Even though this is not something a programmer consciously performs in practice as a concrete, high-level task, it is a precursor to program understanding and other tasks such as bug fixing, adding features, refactoring, etc.  It is implicit and embedded into the developer workflow \citep{wells89}.

An initial study in this direction~\citep{Dyer_2022} investigated how 100K open-source Python projects utilize procedural, object-oriented, functional, and imperative paradigms in their source files and at the project level.  They manually classified a small sample of source files and developed an automated classification script that runs in the Boa language~\citep{boa,boa-website}.  During the manual process, they noted that human raters did not fully agree on what the predominant paradigm was and it was not clear how humans were even judging the predominant paradigm in Python code.  They also did not go beyond simply classifying the paradigm(s) observed.  It is still not understood if the predominant paradigm of Python scripts hinders a developer's comprehension or their ability to debug that code.

Python provides a variety of language features with respect to each paradigm.  For example, object-oriented features such as classes, functional features such as lambdas, procedural features such as function definitions, and imperative features such as names and assignments.  Understanding how developers use the different programming paradigm features available to them and the possible effects of those paradigms on their comprehension is an important topic to study.  It has been observed that the first language taught to students in early courses of computer science affects their comprehension of programming~\citep{Siegfried} and thus developing Python code in a predominant paradigm that is different than their first language could have an effect on learning.  When a developer transitions from an imperative paradigm to a more advanced one like object-oriented or functional, it becomes harder for them to grasp new concepts associated with those features~\citep{Floyd1979}.

Further studies demand the need to understand Python features~\citep{Shrestha20} among the developer community, for example, researchers exploring pattern matching using functions, first-class objects along with pre-existing features~\citep{Kohn2020}, dynamic analysis frameworks supporting Python features like lambda~\citep{Aryaz}, or studies that are using Python packages for analyzing relationship structure using feature definitions, imports, and calls~\citep{Romanov2020}.  Such knowledge can also help guide future researchers investigating code smells in Python~\citep{chen16,rahman19} as they could correlate existing smells to specific paradigms or see if new paradigm-specific smells are occurring. These problems inspired us to look into how developers understand Python's paradigm-specific language features.

  The study we present in this paper explores the question: ``How do developers understand Python's paradigm-specific language features?''
  In particular, we explore the comprehension of paradigm-specific language features through two programming tasks: classification of paradigm and bug localization.
  This helps us better understand how developers read code~\citep{just1980theory}, using an eye tracker~\citep{duchowski2017eye}, that collects time-stamped events on code tokens.
  This methodology has been used within the software engineering community over several years to understand developers' thought processes~\citep{ObaidellahACM2018,Sharafi2015}.
  Eye tracking has provided valuable insights into how developers work and how their thought processes differ based on factors such as expertise or task~\citep{abid2019developer} all within a realistic development environment.
  In particular, the data from this study will help us build new and refine older theories of program comprehension~\citep{iwpc-Storey05} for mixed paradigm languages, which are sorely lacking within the field, and may help us to design more usable programming languages, as well as improve education.
  This study was performed in a realistic development environment (an IDE), with 29 participants of different experience levels in Python.
  All participants were trained to classify Python features by paradigm and asked to complete the two programming tasks.
We find the following key results:

\vspace{-0.3em}
\begin{enumerate}
    \item During classification, developers appear unsure between Functional and Procedural paradigms but can recognize Object-Oriented code.
    \item During classification, the length of the code appears to drive differences in reading behavior more than the predominant paradigm, but the fixation counts and durations are not linearly correlated with the code length.
    \item During classification, developers fixate on many different token types in all paradigms, but they do not only focus on paradigm-relevant token types.
    \item During debugging, the paradigm of the code influences how accurately developers can spot the bug (Procedural code in particular shows the highest accuracy), however, the particular program being debugged has an impact as well.
    \item During debugging, we find higher fixation counts, longer fixation durations, and significantly different vertical and horizontal reading patterns for Functional code. 
\end{enumerate}
\vspace{-0.3em}

In the next section, we provide background on the classification of predominant paradigms in Python code.  In \Cref{sec:related} we discuss some related prior studies.  The studied research questions are provided in \Cref{sec:rqs}.  The study design is given in \Cref{sec:studyDesign} followed by the results in \Cref{sec:results}.  \Cref{sec:discussion} provides a discussion of the results with threats to the validity discussed in \Cref{sec:threats}. Conclusions and future work are presented in \Cref{sec:conclusion}.

\section{Background on Paradigm Classification}
\label{sec:background}

Python is predominantly object-oriented but also supports other paradigms like procedural and functional.  
For example, the Django framework~\citep{django} supports class-based and function-based views.  Some articles claim function-based views are easier to read and understand~\citep{django-views}, but the framework provides both and is moving more toward the class-based paradigm.
Here we briefly describe some of the features Python supports and how we can classify Python code according to those features based largely on the work of \textcite{Dyer_2022}.  We re-use and slightly modify the running example from their work and show it in \Cref{lst:example} and explain how they would classify this code.  Their classification strategy is based on specific features, where they mapped features as functional based on the Python official documentation's how-to guide for functional programming~\citep{func-howto}.  The full feature mapping is shown in \textcite[Tab.~1]{Dyer_2022}.

\begin{lstlisting}[float,label={lst:example},caption={Example Python code with each statement classified into paradigm(s) (based on \cite{Dyer_2022})}]
class MyCounter:                              # func oo $\label{ln:class-def}$
   x = 1                                      #      oo      imp$\label{ln:x}$
   def val(self):                             #      oo
      def wrapper():                          #      oo proc $\label{ln:def-proc}$
         return self.x                        #      oo proc $\label{ln:proc-return}$
      return wrapper                          # func oo $\label{ln:return-fn}$
   def __iter__(self):                        # func oo $\label{ln:iter}$
      self.x = 1                              #      oo
      return self                             #      oo
   def __next__(self):                        # func oo $\label{ln:next}$
      if self.x > 50:                         #      oo
         raise StopIteration                  #      oo
      tmp = self.x                            #      oo
      self.x += 1                             #      oo
      return tmp                              #      oo
print([y for y in MyCounter() if y % 2 == 0]) # func oo proc $\label{ln:class-use}$
\end{lstlisting}

We can classify the overall code in \Cref{lst:example} as object-oriented since the majority of the code implements a class and uses the class (on line~\ref*{ln:class-use}).  Thus every line has an \texttt{oo} classification associated with it.  Procedural programming supports defining functions which are known as procedures.  When we define a function \texttt{def():} we can pass different arguments.  On lines~\ref*{ln:def-proc} and \ref*{ln:proc-return}, we classify them as procedural.  There is also a procedural call to the print function on line~\ref*{ln:class-use}. Since iteration is considered functional, we also classify the class itself and the methods providing iteration support (lines~\ref*{ln:class-def}, \ref*{ln:iter}, \ref*{ln:next}) as functional.  But the \texttt{wrapper} function is being returned in the method, on line~\ref*{ln:return-fn}.  This is an example of a higher-order function and thus we classify line~\ref*{ln:return-fn} as also being functional.

We can already see that several lines are classified as belonging to multiple paradigms.  Most lines in Python are also imperative by nature, either by referencing names or using assignments.  As such, we could potentially mark almost every line as imperative.  Instead, the prior approach states that things should only be marked as imperative if they were not already classified as another paradigm.  Thus, line~\ref*{ln:x} is also marked as imperative because the object-oriented classification of line~\ref*{ln:x} only came about from the block structure of the code and not from the line in isolation.  In general, very few lines wind up being classified as imperative and thus very few files are predominantly imperative, using this prior classification strategy~\citep{Dyer_2022}. In this study, we ignore the imperative features and focus only on procedural, object-oriented, and functional features.  Finally, the last line is also functional as it contains a list comprehension.

Since each line can be classified into more than one paradigm, it is possible for several paradigms to account for more than half the code.  We follow the classification strategy of \cite{Dyer_2022} to indicate when a file is \emph{mixed paradigm}: if the top two paradigms are each present on 2/3 of the lines, or present on 1/2 the lines and within 20\% of each other, or if the most-used paradigm accounts for less than half the lines and the second most-used is within 10\%.  For all other cases, we classify the code by its \emph{predominant paradigm} (the most-occurring paradigm).

The Boa script from \textcite{Dyer_2022} is also used in our study to classify each file.  Their script relies on the classifications from \Cref{lst:example} as well as some heuristics they discovered during their manual classification process (such as the fact that we do classify line~\ref*{ln:x} as imperative, but most other lines are not).  This script is used to classify every source file used in our study and forms our ground truth for the classification and debugging tasks.

\section{Related Work}
\label{sec:related}

We first describe prior work related to Python language feature studies followed by relevant eye-tracking studies done in software engineering.

\subsection{Language Feature Studies in Python}

\citet{hansen-arxiv2013} conducted an experiment on 10 Python programs that varied in notation to determine which variation was harder to understand. They found that experienced participants performed better but not when certain expectations were violated in the code. They did not consider paradigm classification or feature use in their tasks. 
Next, \citet{Peng21} investigated using several different Python features, such as loops, nested classes, inheritance, and keyword arguments.  They built a tool to automatically identify and measure the use of these features in 35 popular projects.  While the features they studied crossed paradigms, they did not focus on how paradigm-specific subsets were used or the impact of those paradigm-specific features on developers. 

Similarly, \citet{yang22} looked at ``complex'' language features, such as reflection, dynamic features, context managers, and generators.
Although the set of features they studied overlaps with several paradigms (object-oriented and functional), their focus was not on the impact of paradigms, but the impact of features.  This work did not perform a user study to determine how those features impact developers, while our study aims to see if paradigms affect developer comprehension of code.
Further studies, such as performed by \citet{keshk23} studied one specific feature in three languages, including Python: the ability to chain method calls together.  They showed the use of method chains differs in Python compared to other languages like Java.  Their study used repository mining techniques and did not investigate what impact the use of chaining might have on developers.

Finally, there has also much research on idiomatic code in Python~\citep{Alexandru,Zhang22,Zhang23,Zhang23b,Farooq21}.  While some research has looked at identifying idioms in Python, other work has attempted to transform existing code to be more idiomatic.  All of these works looked at idioms instead of specific language features, and most only statically analyzed existing repositories instead of performing a user study to measure the potential impact on developers. The study presented in this paper bridges this gap in prior literature by providing objective metrics on what Python developers actually look at while solving tasks.  It follows a developer-centric instead of an artifact-centric approach.

\subsection{Eye-tracking Studies}

Eye tracking has been used for a number of years to explore text comprehension, with wide-ranging results~\citep{Rayner1998}.
Additionally, it has been used for a number of software engineering studies, many of which have been conducted using the Java language~\citep{SHARAFI201579,ObaidellahACM2018}.
In particular, studies have been conducted on various topics such as identifier style~\citep{Binkley:2013}, reading order~\citep{PeitekSA20, busjahn2015eye}, bug fixing~\citep{Kevic2015}, scope highlighting~\citep{Park-ICER23}, code summarization~\citep{Rodeghero2014,abid2019developer}, atoms of confusion~\citep{Oliveira20,dacost23}, and the very first paper on eye tracking in comprehension by \textcite{crosby90}.

We are aware of two eye-tracking studies done on Python and present them next. \citeauthor{Turner:2014:ESA:2578153.2578218} conducted a study on small code snippets in Python~\cite{Turner:2014:ESA:2578153.2578218} that contained bugs. They found no significant difference in bug-finding accuracy or time, however, there was a difference in the rate at which buggy lines were looked at---one difference being that novices had higher fixation durations on buggy lines in Python. 
In recent work, \citeauthor{dacost23} conducted a controlled experiment with 32 novices in Python using an eye tracker~\citep{dacost23} and found that code that had operator precedence took more attempts to solve. In contrast, code with multiple variable assignments had 60\% fewer regressions, indicating more clarity.  
We are unaware of other eye-tracking studies focusing on feature classification and debugging in Python. Our study seeks to bridge this gap by studying the eye movements of developers as they classify and debug in Python. We significantly add more depth to prior analyses by using additional linearity metrics~\citep{busjahn2015eye} and scanpath visualizations via scarf plots~\citep{Alpscarf18}. 

\subsection{Bugginess and Language Paradigms}

Prior research has explored the effect of paradigm (and other language features) on the presence of bugs, with some research finding an increase in bugs in the presence of functional code~\citep{ray14}.
In contrast, later research found fewer associations with bugginess~\citep{berger19}.
This work in contrast looks at a single language and examines how effectively developers can detect bugs in code written in different paradigms within the same language.% \sam{}

\section{Research Questions}
\label{sec:rqs}

  As the main goal of this study is to uncover how paradigm-specific language features impact developer comprehension, we pick two important programming tasks to study.
  The first is paradigm classification, which plays an early role in developer comprehension; here we seek to understand how well developers classify paradigms, and how their reading behavior changes between different paradigms.  We devise the following two research questions for paradigm classification.

\begin{rqs}
\item \label{rq1} How accurately and efficiently do Python developers classify the predominant paradigm?
  To explore this, we use a questionnaire and timing information, to understand ability and efficiency.
\item \label{rq2} How do different predominant paradigms impact developers' reading when classifying for predominant paradigm? For example, do developers look at paradigm-specific tokens? Are developers' reading patterns more or less linear in certain paradigms?
  To explore this, we use eye tracking data, including fixations-on-AOIs (areas of interest), and various linearity metrics.
\end{rqs}

With these results in mind, we want to understand how the paradigm of code affects a particular downstream task: localization of logical errors in code.
This is an important downstream task, and we explore it both using the lenses of efficiency and gaze studied via the following two research questions:

\begin{rqs}
\item \label{rq3} How do different predominant paradigms affect a Python developer’s ability and efficiency to debug logical errors in code?
  To explore this, we use a questionnaire and timing information, to understand ability and efficiency.
\item \label{rq4} How do different predominant paradigms impact developers' reading when debugging logical errors? For example, do developers look at paradigm-specific tokens? Are developers' reading patterns more or less linear in certain paradigms?
  To explore this, we use eye tracking data, including fixations-on-AOIs (areas of interest), and various linearity metrics.
\end{rqs}

In this exploratory study, we hope to start gaining some insight into how code written predominantly in one of the paradigms namely, Python, supports influences the comprehension of that code (\ref*{rq1} and \ref*{rq3}).  Since Python is a multi-paradigm language, it is unclear if developers know what paradigm is being predominantly presented or if the token-level language features help them determine that.  It is also not clear if writing code in one specific paradigm over another leads to lower comprehension of errors occurring in that code.  With \ref*{rq2} and \ref*{rq4}, we also seek to uncover via eye tracking how developers actually read and navigate the code in the different paradigms for classification and debugging tasks and use specific metrics to operationalize the reading behavior. 

\section{Study Design}
\label{sec:studyDesign}

This section discusses our study design, tasks, participants, procedures, and analysis methods. All de-identified study data and processing scripts are available in our replication package~\citep{replication-package}.

\subsection{Terminology}
\label{sec:terminology}

We define the following terms for use in this paper.

\begin{description}[format=\bfseries]
\item[logical bug] A non-syntactic bug, violating the semantics or logic of the task, for example, computing \(x \times x \times x \times 3\) to cube a value instead of computing \(x \times x \times x\) or \(x^{3}\) (see \cref{fig:debugging-cube}).
\item[code paradigm] One of ``functional'', ``object-oriented'', ``procedural'', or ``imperative''.
  This taxonomy, and its definition, was developed previously by \citet{Dyer_2022}.
\end{description}

\subsection{Task Categories}

We divided study tasks into two categories: classification and bug localization.  Each task category had four code snippets with corresponding questions for each. Each code snippet was shown as several variants.  The variant for the classification task was the length of the code snippet in a specific paradigm, whereas the variant for the bug localization task was the program shown in a specific paradigm. 
Each participant was first shown four questions from the classification task category in the functional, mixed, object-oriented (OO), and procedural paradigms with a randomly chosen code length (small, medium, large).  The second set of tasks was from the bug localization category, where the programs (cube, largest, palindrome, factorial) were randomly selected in the various paradigms. Refer to 
\Cref{tab:taskRandomization} for a sequence of how the questions in each task category were assigned to participants. We describe each of the task categories next. 

\begin{table}
    \centering
    \caption{Assigning participants to specific tasks. The bold lines show the sequence for Participant 2.}
    \label{tab:taskRandomization}
    \footnotesize
\begin{tabular}{cllll}
\toprule
& \multicolumn{4}{c}{\textbf{1. Classification Tasks}}\\
\textbf{Participant} & \textbf{A. Functional} & \textbf{B. Mixed} & \textbf{C. OO} & \textbf{D. Procedural}\\
  \cmidrule{2-5}
1        & Small    & Medium   & Large    & Small    \\
2        & \textbf{Medium}   & \textbf{Large}    & \textbf{Small}    & \textbf{Medium}   \\
3        & Large    & Small    & Medium   & Large    \\
\multicolumn{1}{c}{$\vdots$} & \multicolumn{4}{c}{$\vdots$}\\
  29       & Medium   & Large    & Small    & Medium   \\
  \midrule
& \multicolumn{4}{c}{\textbf{2. Bug Localization Tasks}} \\
& \textbf{A. Cube} & \textbf{B. Factorial} & \textbf{C. Largest} & \textbf{D. Palindrome} \\
  \cmidrule{2-5}
  1 & Functional & OO         & Procedural & Mixed \\
  2 & \textbf{OO}         & \textbf{Procedural} & \textbf{Mixed}      & \textbf{Functional} \\
  3 & Procedural & Mixed      & Functional & OO \\
  \multicolumn{1}{c}{$\vdots$} & \multicolumn{4}{c}{$\vdots$}\\
  29 & OO         & Procedural & Mixed      & Functional \\
\bottomrule
\end{tabular}
\end{table}

\paragraph{Classification Task Category.}
The first task category was designed to support answering \ref*{rq1}--\ref*{rq2}.  These questions are classification tasks, where participants are asked to classify the predominant paradigm (Functional, Procedural, or Object-Oriented) in Python code.  We obtained real-world Python source file URLs using Boa's ``2021 Aug/Python'' dataset~\citep{boa}. From that dataset, files were chosen in three different length categories. The lengths were determined using the number of statements in the file (as Boa does not provide line numbers), with a goal of having lengths that are less than a screen of code, about a full screen, and more that would require scrolling. There were three lengths: \textit{Small} (10-15 statements), \textit{Medium} (16-30 statements), and \textit{Large} (31-45 statements). There were a total of four code snippets (Procedural, Object-Oriented, Functional, and Mixed) and each of them was accompanied by a set of survey questions. Participants were asked to classify the code examples based on the predominant paradigm and choose what percent of the code belonged to each paradigm.  An example of the classification task can be seen in \Cref{lst:example}, where each statement is classified into a programming paradigm(s).

\paragraph{Bug Localization Task Category.}
The second task category was designed to support answering \ref*{rq3}--\ref*{rq4}.  These tasks are bug localization tasks to measure how well participants comprehend the given code in a specific paradigm.  Participants were asked to find and explain a bug in the given source code.  The code's first two lines stated the purpose of the code. There were four different variants of programs provided: \textit{Cube} calculates the cube of a number, \textit{Largest} finds the largest number in a list, \textit{Palindrome} checks if the input string is a palindrome, and \textit{Factorial} finds the factorial of a number. 
For each, a single logical bug (a non-syntactic bug, which violates the semantics or logic of the task) was introduced.
We had four variants of each program, one for each paradigm.
Each variant had the same logical bug applied to it, expected the same input, and generated the same output.
An example of a bug localization task is shown in \Cref{fig:debugging-cube}, where the bug was the addition of ``\texttt{* 3}'' when cubing the value.

\begin{figure}[htbp]
  \centering
    \begin{subfigure}{0.55\linewidth}
\begin{lstlisting}
items = [1, 2, 3, 4]

_ = list(map(print, [x * x * x /*\two{*~3}*/ for x in items]))
\end{lstlisting}
    \caption{Functional code}
  \end{subfigure}
  \begin{subfigure}{0.44\linewidth}
\begin{lstlisting}
items = [1, 2, 3, 4]

for i in items:
   print(i * i * i /*\two{*~3}*/)
\end{lstlisting}
    \caption{Mixed-paradigm code}
  \end{subfigure}
  \\\vspace{1em}\par
  \begin{subfigure}{0.55\linewidth}
\begin{lstlisting}
items = [1, 2, 3, 4]

class Cuber:
   def __init__(self, x):
      self.x = x

   def cube(self):
      print(self.x * self.x * self.x /*\two{*~3}*/)

for number in items:
   c = Cuber(number)
   c.cube()
\end{lstlisting}
    \caption{Object-Oriented code}
  \end{subfigure}
  \begin{subfigure}{0.44\linewidth}
\begin{lstlisting}
items = [1, 2, 3, 4]

def cube(x):
   return x * x * x /*\two{*~3}*/

for number in items:
   res = cube(number)
   print(res)
\end{lstlisting}
    \caption{Procedural code}
  \end{subfigure}
  \caption{The \textit{Cube} program for the bug localization task category, shown in all four paradigms. Note that each file had a two-line header describing the task (``Find a logical bug in the code''), and the purpose of the code (``The following code prints the cubed values of a list of numbers'').}
  \label{fig:debugging-cube}
\end{figure}

\subsection{Study Variables}

\paragraph{Independent Variables.}
The common independent variable for the classification and bug localization tasks is the \textit{paradigm} the code was shown in.
The paradigm's treatment in both task categories was one of four types: Functional, Mixed, Object-Oriented, and Procedural.
In addition, the classification task had another independent variable: \textit{code length}---with Small, Medium, and Large treatments.
The bug localization task exchanged code length for a different independent variable: \textit{program}---with Cube, Factorial, Largest Number, and Palindrome treatments.
The study used a within-subjects design for each independent variable (paradigm, code length, and program).
This design allows us to better explore variation caused by the different factors, mitigate learning effects, and improve sampling efficiency (allowing us to maximally utilize limited participant time).

\paragraph{Dependent Variables.}
We analyze several dependent variables (detailed in \Cref{tab:analyzed-variables}) namely correctness, time, confidence, and eye tracking metrics including scan paths~\citep{Sharafi2015}.
The answers to the questionnaire were used to collect correctness, time, and confidence (measured on a five-point Likert-type scale from ``Not confident'' to ``Very confident'').
The use of time is a proxy for task difficulty, and developer performance: faster completion likely indicates that code was easier to understand.
The raw gaze data was used to generate fixations and saccades~\citep{IDT}.
We measured fixation-based and linearity metrics~\citep{busjahn2015eye} such as the amount of time developers spent reading a line, moving up/down in the code, re-reading code, and navigation (denoted by saccade length) using the eye-tracking data.
Fixations indicate focus and attention, and fixation-based metrics~\citep{IDT} are widely used in eye-tracking studies to analyze code reading behavior and attention~\citep{sharif2010eye, busjahn2014eye, busjahn2015eye, abid2019developer, beelders2016influence}.
Such metrics are associated with reading behaviors which are influenced by program design, and may be influenced by paradigm-specific tokens or constructs.
This paper uses fixation count, fixation duration, and normalized mean fixation duration.
For normalization, each fixation duration is divided by the number of characters of the fixated token, then the mean is taken.
This allows us to consider the amount of time spent on paradigm-specific tokens while avoiding issues caused by token length.
In addition to the fixation metrics, we use the linearity metrics designed by \citet{busjahn2015eye}.
The linearity metrics show vertical and horizontal reading patterns, saccade length (length between two consecutive fixations), and regressions.
We used the variables collected from the questionnaire to answer \ref*{rq1} and \ref*{rq3}, and the eye-tracking data variables to answer \ref*{rq2} and \ref*{rq4}.
These reading patterns should vary by paradigm, as different paradigms follow more or less linear patterns of organization, or exhibit different patterns of information locality.

\begin{table}[ht]
    \centering
    \caption{Dependent variables and their measures, mapped to research questions.}
    \label{tab:analyzed-variables}
    \vspace{-0.5em}
\renewcommand{\tabcolsep}{0.25em}
\begin{tabularx}{0.95\linewidth}{llX}
  \toprule
  \textbf{RQs} & \textbf{Variable}& \textbf{Measure} \\
  \midrule
  1, 3 & correctness of response & choice of paradigm \\
  1, 3 & confidence of response & 5 point Likert-type \\
  1, 3 & time to completion & time in seconds \\
  \midrule
  2, 4 & fixation count & count (frequency) \\
  2, 4 & fixation duration & time in milliseconds (ms) \\
  2, 4 & vertical next text & forward saccades (\%) on same or one line down \\ 
  2, 4 & vertical later text & forward saccades (\%) on same or any number of lines down \\
  2, 4 & horizontal later text & forward saccades (\%) on same line \\
  2, 4 & line regression rate & backward saccades (\%) on same line\\
  2, 4 & saccade length & mean Euclidean distance between consecutive fixations\\
  2, 4 & scanpaths & scarf plots  \\
  \bottomrule
\end{tabularx}
\end{table}

\subsection{Participants and Grouping}

We recruited participants 19 years of age or older from the University of Nebraska--Lincoln, requiring self-reported prior Python experience.
This gave us a total of 31 participants, but two had to be excluded: one for poor calibration, and one for having no prior Python experience.
We analyzed the data from the remaining 29 participants.
The majority of participants were students from the University of Nebraska--Lincoln. The majority of the students were from a computer science background and others were from biological systems engineering, statistics, computer engineering, and other related engineering fields. Because of varying backgrounds, we cannot verify programming language background or history of the participants. The participants were not provided any course incentives but were compensated with a \$25 virtual Visa card.

Each participant was provided with twelve questions in total, six in each task category.
Participants provided their responses to each question by typing their answers into a web form using a provided tablet.
Classification tasks were assigned to ensure that roughly one-third of participants saw a small+functional code, one-third saw medium+functional, and one-third saw large+functional.  This was done for each of the four paradigms.  Similarly, for the bug localization category, 
tasks were assigned to ensure roughly equal coverage among each program and paradigm.
\Cref{tab:taskRandomization} shows an example grouping for four participants.

\subsection{Eye-Tracking Apparatus and Environment}

To track the participants' eyes, we used the Tobii TX300 eye tracker running at 60Hz frequency.  Participants used Eclipse as the integrated development environment (IDE) for viewing all code snippets. The iTrace infrastructure~\citep{p-guarnera-iTrace-2018,i-Trace2020} was used to map the raw eye coordinates to a line and column in the file as the participant looked at the source code while performing tasks. The iTrace-Eclipse plugin provides an output of two XML files containing time-stamped session logs. The data is then passed through the Identification by Dispersion Threshold (IDT) fixation filter~\citep{IDT} with the dispersion threshold parameter set to 125 pixels and the duration threshold set to 10 milliseconds using iTrace-Toolkit~\citep{iTraceToolkit} to generate the fixations over source code. At the time of this writing, iTrace does not natively support mapping of eye gaze to source tokens for Python. We wrote a Python script to generate the mapping from gazes to tokens using the line and column information generated in the iTrace databases.
Other than mapping tokens, we performed no other post-processing of eye-tracking data.

\vspace{-0.2em}
\subsection{Study Procedure}

The study was conducted in a dedicated eye-tracking lab. Participants were allocated up to two hours to complete the entire study. After signing a consent form, participants filled out the pre-questionnaire at the start with some demographic information. They were then given the training task. Next, each participant had to be calibrated with the eye tracker, where they learned how to sit and position themselves in front of the monitor properly. Once the calibration was done, the participant was ready to perform tasks.  For each task, the participants were asked to read code in Eclipse and answer questions about the code.  Each task was allocated a maximum of 10 minutes for participants to complete.  After each task category, a debriefing was done as part of the protocol. 
Finally, participants filled out a post-questionnaire where they self-rated their Python expertise levels. We did this last to avoid any stereotype threats~\citep{p-spencer1999stereotype} that might affect study performance.

\subsubsection{Participant Training}

In order to perform the study tasks, we first trained all participants.
  This process was started by first showing the participants a reference sheet that consisted of mock code snippets for the classification and debugging tasks.
  These mock snippets were built from \mbox{\citet[fig.~1]{Dyer_2022}}, which explained how the paradigms use Python features, by classifying each statement.
They were also provided with the feature classification table as shown in \citet[Tab.~1]{Dyer_2022}, which was available at any time during the classification portion of the study.
Here, during training, we introduced the concept of predominant paradigm, wherein they saw how the training code has a predominant paradigm of object-oriented since all statements are part of a \texttt{class MyNumbers} for instance.
While understanding every feature-wise classification per statement, they were also made aware of how one statement can belong to multiple paradigms.
There was also a bug localization training task where we seeded a fault by changing from \texttt{+=} to \texttt{-=} (see replication package~\cite{replication-package}).
    By training participants on how to classify features, they should pay more attention to paradigm-specific features, thus impacting downstream measures.

\subsubsection{Analysis}

To analyze the effects of the presented paradigm on correctness, we use Cochran's \(Q\)~\citep{cochran50:_compar_percen_match_sampl}, with individual subjects as blocks, and the presented paradigm (or code length, or program) as treatments.
To quantify the effect of treatment overall, we use the \(\eta^{2}_{Q}\) as proposed by \citet{berry07:_alt_measure_effect_size_Q}.
To follow up on the omnibus \(Q\), we use McNemar's \(\chi^{2}\) test~\citep{mcnemar47:_note_sampl_error_diff_corr_prop} and Bonferroni's correction to control the family-wise error rate.
Finally, to analyze relationships between confidence and the presented paradigm, length, or program, we use Fisher's Exact Test~\citep{fisher22:_inter_chisq_from_contin_tables_calcul_p}, because we have small cell counts (\(< 5\)) which would otherwise cause problems for the usage of Pearson's \(\chi^{2}\).

To analyze the effects of task and presented paradigm on time-on-task and various fixation and linearity metrics, we use within-groups factorial ANOVAs as omnibus tests, \citep[p. 210]{boslaugh13:_statis_nutsh}.
For these analyses, we report degrees of freedom, the \(F\) statistic, and \(p\) values, as well as the effect sizes \(\eta^{2}\), \(\eta^{2}_{p}\) (partial), and Cohen's \(f\).
We then follow-up the omnibus ANOVAs with pairwise comparisons performed using Tukey's HSD~\citep{tukey49:_compar_indiv_means_analy_varian}, a single-step multiple comparisons test that produces adjusted \(p\)-values.

Finally, to determine if paradigm-specific tokens are fixated on longer, we use a simple \(t\)-test between count of fixations in and not in areas of interest (AOI) for each task/paradigm pair.
We follow this up with Bonferonni's correction to avoid the family-wise error rate problem.
A listing of paradigm-specific tokens, based on \cite{Dyer_2022} is shown in \Cref{tab:paradigm-tokens}. 

\begin{table}[htbp]
  \centering
  \caption{Paradigm-Specific Tokens}
  \label{tab:paradigm-tokens}
\begin{tabular}{cccc}
\toprule
\textbf{Functional} & \textbf{OO} & \textbf{Procedural} & \textbf{Mixed} \\
Lambda & Attribute & arg & Attribute \\
Call & ExceptHandler & Call & arg \\
GeneratorExp & Call & Return & Lambda \\
ListComp & Raise & FunctionDef & ExceptHandler \\
 & With &  & Call \\
 & FunctionDef &  & Return \\
 & ClassDef &  & Raise \\
 &  &  & With \\
 &  &  & FunctionDef \\
 &  &  & GeneratorExp \\
 &  &  & ClassDef \\
 &  &  & ListComp \\
\bottomrule
\end{tabular}
\end{table}

Statistical analyses are performed using GNU R~\citep{r-lang}, with the \texttt{effectsize} package~\citep{ben-shachar20:_effectsize} to calculate \(\eta^{2}\), and the \texttt{nonpar} package~\citep{sweet20:_nonpar} to calculate Cochran's \(Q\).
Statistical visualizations are produced using \texttt{matplotlib}~\citep{hunter07:_matpl} and \texttt{seaborn}~\citep{waskom21}, with data handling provided by \texttt{pandas}~\citep{pandas20}.
Specific version information and environment definition are in our replication package~\citep{replication-package}.

In order to analyze scanpaths (how one reads i.e., moves from one fixation to next) for RQ2 and RQ4, we take advantage of the scarf plot visualization~\citep{Alpscarf18}. The scarf plot shows eye transitions between different parts of the code snippets and are mainly used to visualize gaze transitions within areas of interest in time. 
The x-axis in a scarf plot represents the task timeline showing duration of how long was spent viewing the token type in the code snippet. The y-axis shows each participant in the group on the timeline. The correctness of each participant is also reported near the y-axis to allow to look for trends in eye movement scanpaths between correct vs. incorrect answers.  
See Figure \ref{fig:classify_f_s} for an example of a scarf plot.  

\section{Results}
\label{sec:results}

\subsection{Participant Demographics}
\label{sec:part-demogr}

To understand participant expertise in Python, we ask for a self-report in a post-questionnaire to avoid priming and stereotype effects~\citep{p-steele1995stereotype,p-spencer1999stereotype, p-shapiro2007stereotype}.
We observe that the majority rated themselves as having intermediate Python skills (69\%), few had Expert-level skills (10\%), with the remainder being Novice (21\%).
58\% of participants had two or less years of experience, while 31\% had 2-5 years of experience.  Only 10\% of participants have five or more years of experience, which matches the number that self-reported as experts.
These and other questions (such as how developers view coding in Python and how frequently they code in Python) are summarized in \Cref{tab:demographics}.

\begin{table}[htbp]
 \centering
 \caption{Demographics for the 29 participants (after 2 excluded)}
 \label{tab:demographics}
 \vspace{-0.5em}
 \begin{subtable}[t]{0.5\linewidth}
   \centering
   \caption{``How would you rate your programming in Python?''}%Self-reported rating for Python programming skills}
   \label{tab:ExpertLevel}
   \small
\begin{tabular}{lr}
\toprule
\textbf{Novice} & 21\% \\
\textbf{Intermediate} & 69\% \\
\textbf{Expert} & 10\% \\
\bottomrule
\end{tabular}
 \end{subtable}
 \begin{subtable}[t]{0.49\linewidth}
   \centering
   \caption{``How often do you code using Python?''}
   \label{tab:Confidencelevel}
   \small
\begin{tabular}{lr}
\toprule
\textbf{Very Frequently} & 17\% \\
\textbf{Frequently} & 34\% \\
\midrule
\textbf{Occasionally} & 41\% \\
\textbf{Very Rarely} & 7\% \\
\bottomrule
\end{tabular}
 \end{subtable}
 \vspace{-0.5em}
 \newline
 \begin{subtable}[t]{0.5\linewidth}
   \centering
   \caption{``How long have you been coding using Python?''}
   \label{tab:YearsofExperiencewithPython}
   \small
\begin{tabular}{lr}
\toprule
\textbf{Less than or equal to 1 year} & 24\% \\
\textbf{Between 1-2 years} & 34\% \\
\midrule
\textbf{Between 2-3 years} & 14\% \\
\textbf{Between 3-5 years} & 17\% \\
\midrule
\textbf{Greater than or equal to 5 years} & 10\% \\
\bottomrule
\end{tabular}
 \end{subtable}
 \begin{subtable}[t]{0.49\linewidth}
   \centering
   \caption{``How important is it for you to code in Python?''}
   \label{tab:ImportanceofPython}
   \small
\begin{tabular}{lr}
\toprule
\textbf{Very Important} & 31\% \\
\textbf{Important} & 31\% \\
\midrule
\textbf{Moderately Important} & 28\% \\
\textbf{Slightly Important} & 3\% \\
\midrule
\textbf{Not Important} & 7\% \\
\bottomrule
\end{tabular}
 \end{subtable}
\end{table}

\subsection{\ref*{rq1} Results: Accuracy and Efficiency of Paradigm Classification}
\label{sec:rq1-results}

\subsubsection{Classification Accuracy}
\label{sec:rq1-class-accuracy}

We show the accuracy of judgment, broken down by both paradigm and code size, in \Cref{tab:classification-crosstab}.
Of note, the presented paradigm does influence how effectively participants were able to classify code (\statistic{Q(2)}{9.5}, \pvalue{0.00865}, \statistic{\eta^{2}}{0.163793}).
In particular, we find that there is no difference between Object-Oriented code and Functional code in terms of the ability to classify (\chisquare{1}[1.125], \pvalue[adj]{0.86653310}), nor between Functional code and Procedural code (\chisquare{1}[2.08333], \pvalue[adj]{0.44674402}), but there is a difference between Object-Oriented code and Procedural code (\chisquare{1}[6.75], \pvalue[adj]{0.02812431}).
This comparison shows that participants were able to correctly classify Object-Oriented code more often (86\% of the time) than Procedural code (52\% of the time).

\begin{table}[ht]
  \centering
  \caption{Accuracy of paradigm decision. Participants were asked: ``What do you consider the predominant programming paradigm for the code you read?'' The presented paradigm is shown in rows and code length is in columns. Highlighted cells represent stimuli categories that showed a statistical difference.}
  \label{tab:classification-crosstab}
  \small
\begin{tabular}{lrrrr}
\toprule
 & \textbf{Small} & \textbf{Medium} & \textbf{Large} & \textbf{Overall} \\
\cmidrule{2-4}
\textbf{Functional} & 8 / 10 & 5 / 10 & 8 / 9 & 72.41\% \\
\textbf{Object-Oriented} & 9 / 10 & 7 / 9 & 9 / 10 & \highlight{86.21\%} \\
\textbf{Procedural} & 6 / 10 & 5 / 10 & 4 / 9 & \highlight{51.72\%} \\
\midrule
\textbf{Overall} & 79.31\% & 58.62\% & 72.41\% &  \\
\bottomrule
\end{tabular}
\end{table}

To further explore this difference, we look at \Cref{tab:percentage-paradigms} which shows how participants binned code into paradigms.
This shows us that Object-Oriented samples (row two) were consistently rated as being \(> 75\%\) Object-Oriented, with participants consistently considering it as \(\leq 50\%\) Functional or Procedural.
Participants, however, were more confused by Functional (row one) and Procedural (row three) samples.
Functional code was thought to be at least 75\% Functional by most participants, but almost a third thought it was at least 75\% Procedural; this pattern is flipped for Procedural.

\begin{finding}{}{fd:confusion}
    When classifying the predominant paradigm, there appears to be confusion in labeling Functional and Procedural paradigms, while classifiers are accurate on Object-Oriented.
\end{finding}

\begin{table}[t]
  \centering
  \caption{Classification of code. Participants were asked: ``What percentage of the code falls under the following paradigms? The percentages do not have to add up to 100\%.'' Rows represent answers for each presented paradigm. Highlighted cells are high values for each presented paradigm/responded paradigm pair.}
  \label{tab:percentage-paradigms}
\newcommand{\oldtabcolseprqpercents}{\tabcolsep}
\renewcommand{\tabcolsep}{3pt}
\rowcolors{3}{gray!10}{white}
\begin{tabular}{lrrrrrrrrrrrr}
\toprule
 & \multicolumn{4}{c}{\textbf{Functional}} & \multicolumn{4}{c}{\textbf{Object-Oriented}} & \multicolumn{4}{c}{\textbf{Procedural}} \\
 & \textbf{\textless{}25\%} & \textbf{25--50\%} & \textbf{51--75\%} & \textbf{\textgreater{}75\%} & \textbf{\textless{}25\%} & \textbf{25--50\%} & \textbf{51--75\%} & \textbf{\textgreater{}75\%} & \textbf{\textless{}25\%} & \textbf{25--50\%} & \textbf{51--75\%} & \textbf{\textgreater{}75\%} \\
\cmidrule(r){2-5}\cmidrule(lr){6-9}\cmidrule(l){10-13}
\textbf{Func.} & 3 & 4 & 5 & \highlight{17} & \highlight{16} & 12 & 1 & 0 & 8 & \highlight{9} & 3 & \highlight{9} \\
\textbf{OO} & \highlight{13} & 9 & 7 & 0 & 1 & 3 & 5 & \highlight{20} & 8 & \highlight{9} & 8 & 4 \\
\textbf{Proc.} & 6 & \highlight{9} & 5 & \highlight{9} & \highlight{17} & 6 & 3 & 3 & 3 & 6 & 6 & \highlight{14} \\
\textbf{Mixed} & \highlight{10} & \highlight{10} & 5 & 4 & 5 & 8 & \highlight{10} & 6 & 7 & 6 & 3 & \highlight{13} \\
\bottomrule
\end{tabular}\renewcommand{\tabcolsep}{\oldtabcolseprqpercents}
\end{table}

\begin{table}[ht]
  \centering
  \caption{Do participants perceive multiple paradigms? Participants were asked ``Do you think the code has more than one predominant paradigm?'' Highlighted cells represent high values for each column.}
  \label{tab:code-is-mixed}
\newcommand{\oldtabcolseprqismixedpivot}{\tabcolsep}
\renewcommand{\tabcolsep}{4pt}
\begin{tabular}{lcccccccccccc}
\toprule
 & \multicolumn{3}{c}{\textbf{Functional}} & \multicolumn{3}{c}{\textbf{Mixed}} & \multicolumn{3}{c}{\textbf{Object-Oriented}} & \multicolumn{3}{c}{\textbf{Procedural}} \\
 & \textbf{Small} & \textbf{Med} & \textbf{Large} & \textbf{Small} & \textbf{Med} & \textbf{Large} & \textbf{Small} & \textbf{Med} & \textbf{Large} & \textbf{Small} & \textbf{Med} & \textbf{Large} \\
\cmidrule(r){2-4}\cmidrule(lr){5-7}\cmidrule(lr){8-10}\cmidrule(l){11-13}
\textbf{No} & \highlight{6} & \highlight{4} & 4 & 2 & 2 & 0 & 3 & 4 & 1 & 3 & \highlight{5} & 0 \\
\textbf{Maybe} & 1 & 2 & 0 & 0 & 2 & 2 & 1 & 0 & 0 & 0 & 2 & 2 \\
\textbf{Yes} & 3 & \highlight{4} & \highlight{5} & \highlight{7} & \highlight{6} & \highlight{8} & \highlight{6} & \highlight{5} & \highlight{9} & \highlight{7} & 3 & \highlight{7} \\
\bottomrule
\end{tabular}\renewcommand{\tabcolsep}{\oldtabcolseprqismixedpivot}
\end{table}

We then consider how often developers believe a code sample contains multiple predominant paradigms.
In \Cref{tab:code-is-mixed} we show this for each paradigm/code length pair.
We see that most participants think the code has multiple \emph{predominant} paradigms.
In the cases of Functional and Procedural, this may help to explain some of the confusion we see in the medium-length Procedural code.
In particular, it shows that while participants did not think that it definitely has multiple predominant paradigms, their understanding of the difference between the two may be flawed.

With respect to the confidence ratings, We find that there does not appear to be a relationship between confidence and the presented paradigm (\pvalue{0.5742}).
There also does not appear to be a relationship between correctness and confidence (\pvalue{0.3132}).
These both make sense, as we see that in 88\% of cases, participants were confident in their answers.

\begin{finding}{}{fd:paradigm}
    Participants felt very confident in their ability to classify the predominant paradigm, regardless of whether they were correct or not.
\end{finding}

\subsubsection{Classification Efficiency}
\label{sec:rq1-class-efficiency}

\begin{figure}[ht]
    \centering
    \includegraphics[width=0.8\linewidth,page=1]{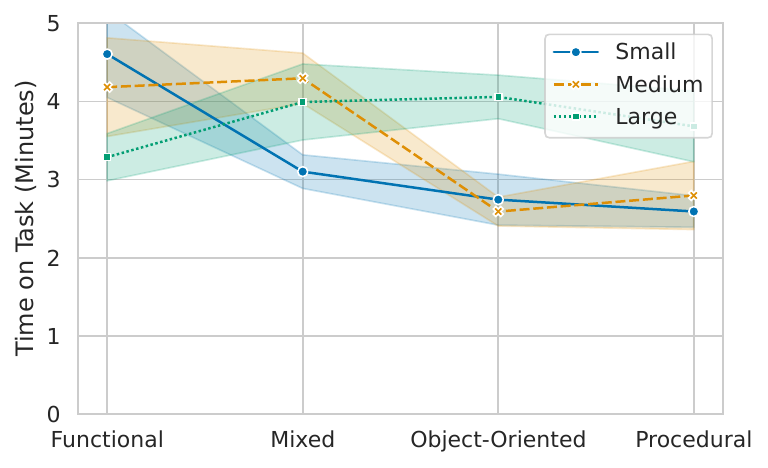}
    \caption{Average time on task, per paradigm, for classification, including size.}
    \label{fig:classification-time-on-task}
\end{figure}

To determine whether or not paradigm and task size have an effect on efficiency (time on task), we ran a within-groups \(4 \times 3\) ANOVA, a summary of which is shown in \Cref{fig:classification-time-on-task}.
These results show that there is a significant difference between presented paradigms (\Fmeasure{3}{104}[4.93], \pvalue{0.0031}, \statistic{\eta^{2}_{p}}{0.12}, \statistic{f}{0.38}) as well as a significant interaction between presented paradigm and code size (\Fmeasure{6}{104}[3.26], \pvalue{0.0056}, \statistic{\eta^{2}_{p}}{0.16}, \statistic{f}{0.43}).
Post-hoc comparison of the presented paradigm effect shows a difference between Functional and Object-Oriented (\pvalue{0.0324}) and Functional and Procedural (\pvalue{0.0087}); in both cases, the Functional code took longer to classify.

The interaction between paradigm and task size is particularly interesting, though not easy to characterize.
When considering the differences highlighted in \Cref{fig:classification-time-on-task}, it appears that the assumed pattern of time-to-completion increasing as code samples became larger does not hold.
In fact, the only differences that we see in the interaction are between the Functional Small and Object-Oriented Small (\pvalue{0.046}), Functional Small and Procedural Small (\pvalue{0.020}), and Functional Small and Object-Oriented Medium (\pvalue{0.027}).
That is, while we see an interaction, it is only a few particular pairs that we see this on, all of which involve the task that apparently took the longest: Functional Small.

\begin{finding}{}{fd:classification-slower}
    When classifying code, developers take more time when presented with predominantly functional code.
    In general, however, longer code does not take longer to classify, though the Functional Small took longest.
\end{finding}

\subsection{\ref*{rq2} Results: Language Feature(s) Gazed Upon During Paradigm Classification}
\label{sec:rq2-results}

We analyze the participants' gazes using both fixation and linearity-based metrics (\Cref{tab:analyzed-variables}).

\subsubsection{Classification Fixation Metrics}
\label{sec:rq2-class-fixation}

To determine which specific Python token-level language feature(s) participants look at when classifying predominant paradigms, we first looked into the eye-tracking fixation metrics (summarized in \Cref{tab:fixation-metrics-classification}) and how they differ between the different paradigms and code lengths in the classification task. We ran a within-groups \(4 \times 3\) ANOVA for fixation count (shown in \Cref{tab:anova-classification-fix-count}), fixation duration, and normalized mean fixation duration. The ANOVAs were conducted on the influence of paradigm and code length on the fixation metrics.
For fixation count, all effects were statistically significant except for paradigm. We find that the code length has a significant effect on fixation count (\Fmeasure{2}{104}[6.20], \pvalue{0.0029}, \statistic{\eta^{2}_{p}}{0.11}, \statistic{f}{0.35}), and that the interaction effect between the paradigm and code length was significant as well (\Fmeasure{6}{104}[2.53], \pvalue{0.0251}, \statistic{\eta^{2}_{p}}{0.13}, \statistic{f}{0.38}). Based on the pairwise comparisons within paradigms and code length (available in the replication package~\cite{replication-package}), it is likely that the code length affects the fixation count more than the paradigm. However, this relationship is not as expected: more fixations did not occur on longer code.

Additionally, we found that while our participants attended to paradigm tokens, it was less frequently than non-paradigm tokens.
The results of a \(t\)-test and mean fixation counts on paradigm/non-paradigm tokens are shown in \Cref{tab:fixation-counts-classification}.

\begin{table}[htbp]
  \centering
  \caption{Fixation counts on paradigm and non-paradigm tokens for code classification.}
  \label{tab:fixation-counts-classification}
  \renewcommand{\tabcolsep}{0.25em}
\begin{tabular}{lrrrrll}
  \toprule
\textbf{Paradigm} & \textbf{Paradigm Tokens} & \textbf{Non-Paradigm Tokens} & \textbf{t} & \textbf{df} & \textbf{p} & \textbf{p (adj)} \\ 
  \midrule
Functional & 19.07 & 94.45 & -6.78 & 34.28 & $<$ .001 & $<$ .001 \\ 
  Mixed & 30.72 & 87.66 & -6.88 & 55.97 & $<$ .001 & $<$ .001 \\ 
  Object-Oriented & 40.55 & 75.76 & -3.63 & 39.21 & $<$ .001 & $<$ .001 \\ 
  Procedural & 23.29 & 93.90 & -5.53 & 32.96 & $<$ .001 & $<$ .001 \\ 
   \bottomrule
\end{tabular}
\end{table}

We do not see any significant effect on fixation duration from paradigms or code length. However, the results of ANOVA for normalized mean fixation duration show significant interactions between the paradigm and code length (\Fmeasure{6}{104}[3.770], \pvalue{0.0019}, \statistic{\eta^{2}_{p}}{0.18}, \statistic{f}{0.47}) and a significant difference between different code lengths (\Fmeasure{2}{104}[5.986], \pvalue{0.0034}, \statistic{\eta^{2}_{p}}{0.10}, \statistic{f}{0.34}).
These results indicate that code length contributed to more of the difference in normalized mean fixation duration than paradigm, similar to what we see in fixation count.

\begin{table}[htbp]
  \centering
  \caption{High-level summary of fixation metrics for classification tasks (ms is milliseconds).}
  \label{tab:fixation-metrics-classification}
\begin{tabular}{lrr}
\toprule
 & \textbf{mean} & \textbf{std} \\
\textbf{Total Duration (s)} & 210.34 & 82.26 \\
\textbf{Total Duration (m)} & 3.51 & 1.37 \\
\textbf{Fixation Count} & 115.82 & 64.18 \\
\textbf{Fixation Duration (ms)} & 103,569.76 & 45,914.62 \\
\textbf{Mean Fixation Duration (ms)} & 1,013.35 & 456.67 \\
\textbf{Normalized Fixation Duration (ms)} & 27,365.56 & 16,286.45 \\
\textbf{Normalized Mean Fixation Duration (ms)} & 259.49 & 131.58 \\
\textbf{Total Fixations} & 115.82 & 64.18 \\
\textbf{Vertical Next Text (\%)} & 0.31 & 0.09 \\
\textbf{Vertical Later Text (\%)} & 0.59 & 0.06 \\
\textbf{Horizontal Later Text (\%)} & 0.09 & 0.04 \\
\textbf{Regression Rate (\%)} & 0.47 & 0.03 \\
\textbf{Saccade Length (px)} & 167.83 & 30.55 \\
\bottomrule
\end{tabular}
\end{table}

\begin{table}[htbp]
  \centering
  \caption{ANOVA Results for Fixation Counts on Classification Tasks}
  \label{tab:anova-classification-fix-count}
\begin{tabular}{lrrrrr}
  \toprule
 & \textbf{Df} & \textbf{Sum Sq} & \textbf{Mean Sq} & \textbf{F value} & \textbf{Pr($>$F)} \\ 
  \midrule
\textbf{paradigm        } & 3 & 584.58 & 194.86 & 0.05 & 0.9833 \\ 
  \textbf{variant         } & 2 & 44567.63 & 22283.82 & 6.20 & 0.0029 \\ 
  \textbf{paradigm:variant} & 6 & 54585.55 & 9097.59 & 2.53 & 0.0251 \\ 
  \textbf{Residuals       } & 104 & 373905.43 & 3595.24 &  &  \\ 
   \bottomrule
\end{tabular}
\end{table}

\begin{finding}{}{fd:paradigm-et-fixations}
Code length, not the paradigm, affects the fixation differences when participants classify code.
\end{finding}

\subsubsection{Visualization of Classification Fixation Metrics}
\label{sec:rq2-fixation-vis}

\begin{figure}[ht]
  \centering
  \includegraphics[width=\textwidth]{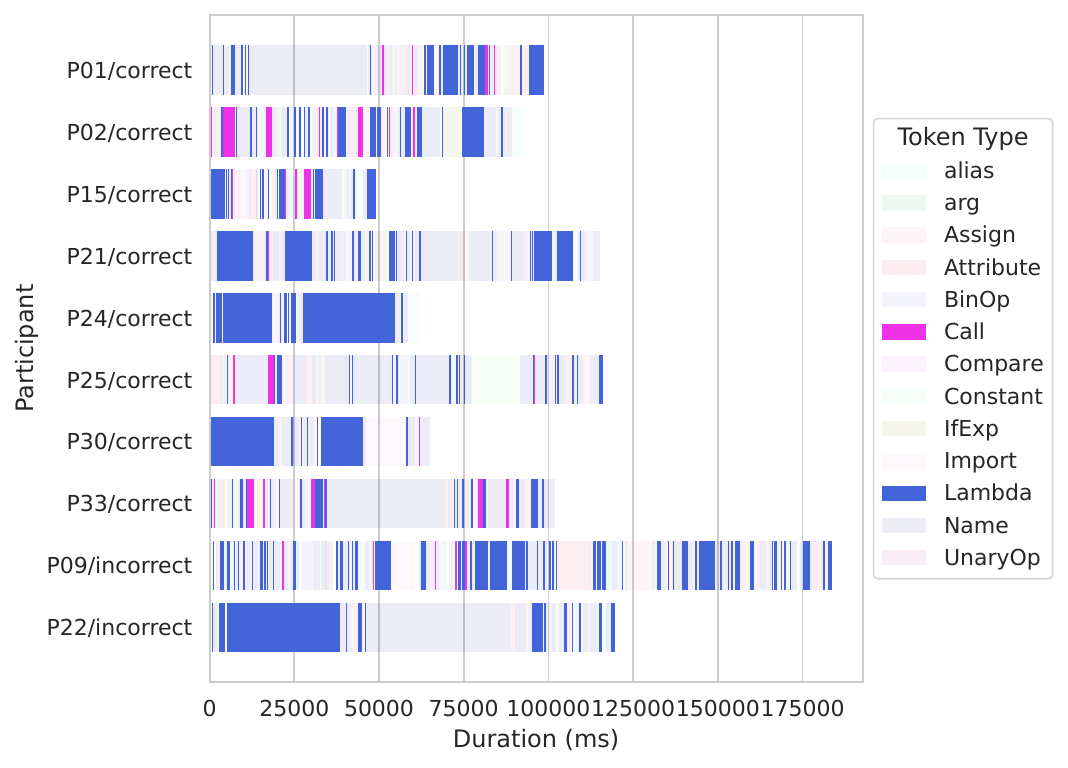}
  \caption{Scarf plots showing fixations over tokens for code classification task, with participants labeled on the \(y\)-axis (participant number and correctness) and time (in \unit{ms}) on the \(x\)-axis: Functional paradigm, Small code length}
  \label{fig:classify_f_s}
\end{figure}

\begin{figure}[ht]
  \centering
  \includegraphics[width=\textwidth]{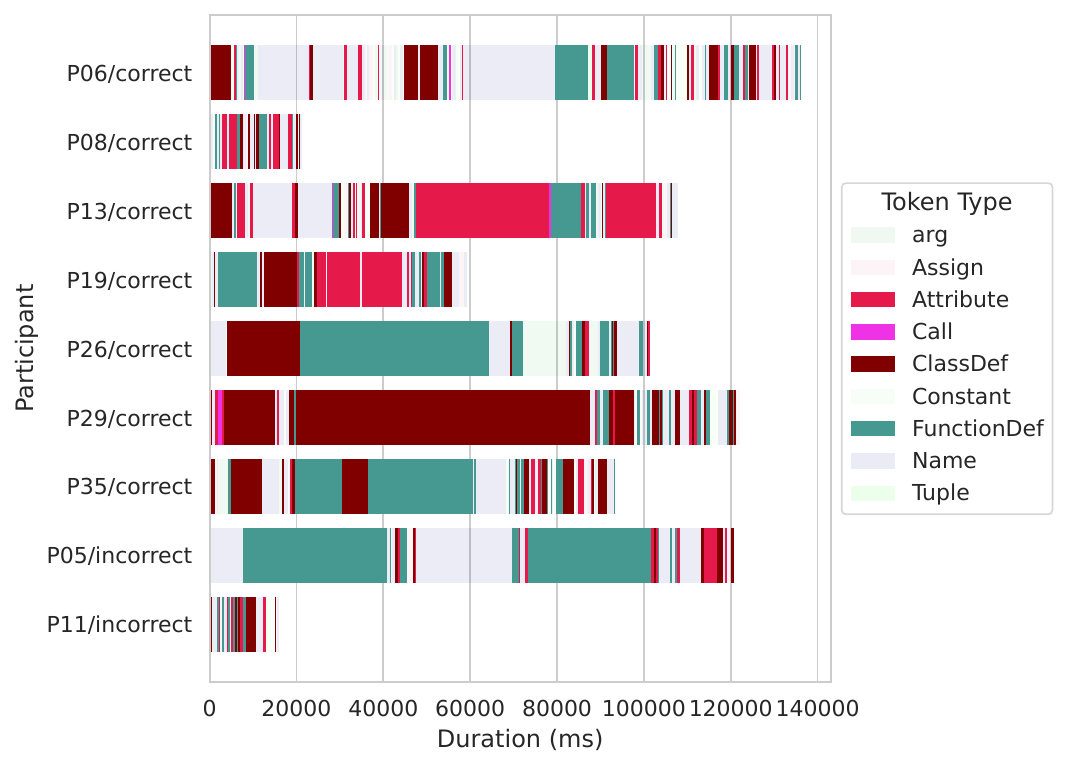}
  \caption{Scarf plots showing fixations over tokens for code classification task, with participants labeled on the \(y\)-axis (participant number and correctness) and time (in \unit{ms}) on the \(x\)-axis: Object-Oriented paradigm, Medium code length}
  \label{fig:classify_oo_m}
\end{figure}

\begin{figure}[ht]
  \centering
  \includegraphics[width=\textwidth]{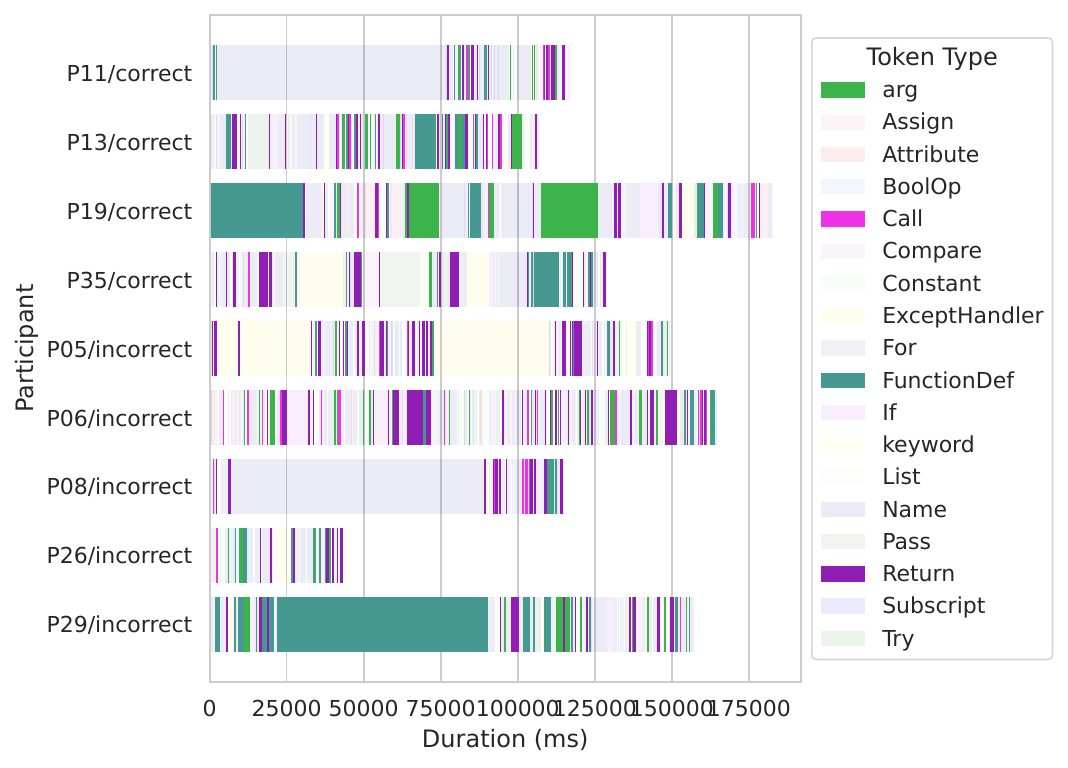}
  \caption{Scarf plots showing fixations over tokens for code classification task, with participants labeled on the \(y\)-axis (participant number and correctness) and time (in \unit{ms}) on the \(x\)-axis: Procedural paradigm, Large code length}
  \label{fig:classify_p_l}
\end{figure}

\begin{figure}[ht]
  \centering
  \includegraphics[width=\textwidth]{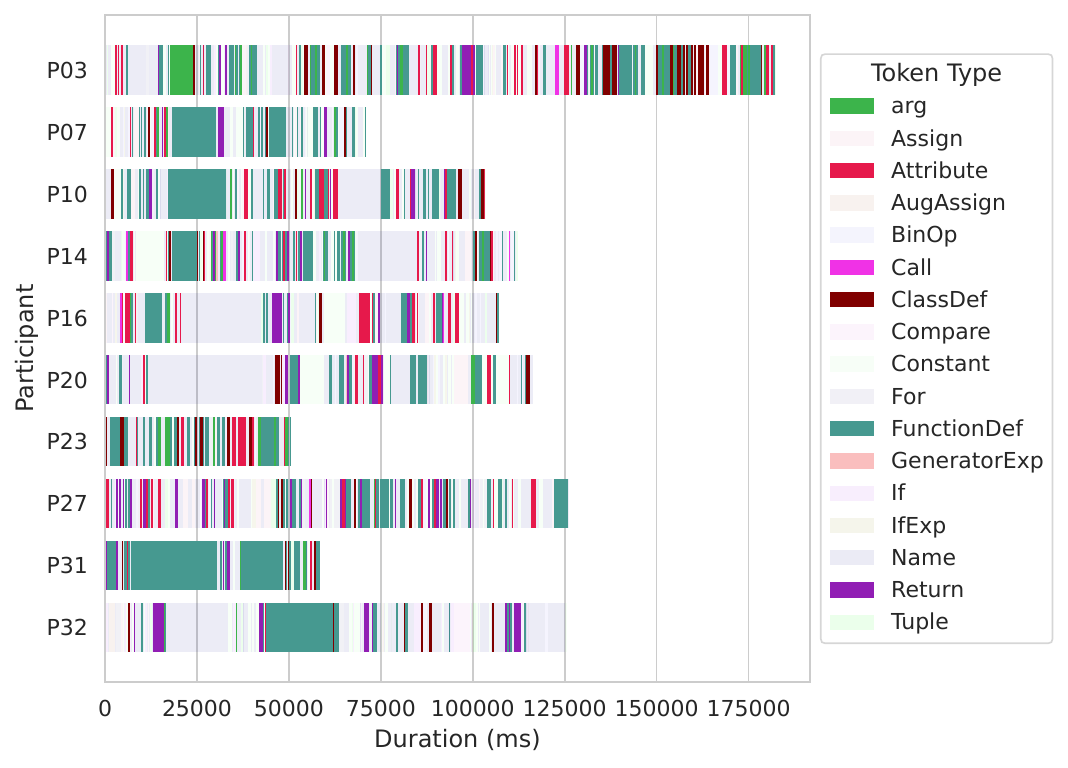}
  \caption{Scarf plots showing fixations over tokens for code classification task, with participants labeled on the \(y\)-axis (participant number and correctness) and time (in \unit{ms}) on the \(x\)-axis: Mixed paradigm, Large code length}
  \label{fig:classify_m_l}
\end{figure}

To visualize the differences in fixations, we created scarf plots showing the fixation duration over different token types (related to language features). We show four different scarf plots, each representative of the scarf plots of one paradigm, in \Cref{fig:classify_f_s,fig:classify_oo_m,fig:classify_p_l,fig:classify_m_l}. All twelve scarf plots are included in the replication package~\citep{replication-package}.
The scarf plots' legends show the token types found in the corresponding code. Token types relevant to the paradigm are specified with vivid colors, whereas the colors of the token types not relevant to the paradigm are faded.

An overview of the scarf plots shows that most participants fixate on relevant token types while reading the code, but they do not show an obvious pattern of (in)correctness being related to fixations on relevant tokens. It can be seen in \Cref{fig:classify_f_s,fig:classify_oo_m,fig:classify_p_l} that the participants who answered incorrectly did spend some time fixating on the paradigm-relevant tokens.

We observe that participants fixated more frequently on \texttt{Lambda} tokens, both in the Small (\Cref{fig:classify_f_s}) and Large code lengths (see the replication~\citep{replication-package}) of the Functional paradigm. For the Object-Oriented paradigm, we observe more frequent fixations on the \texttt{FunctionDef} token type for all different code lengths and prominence of fixations on \texttt{ClassDef} in the Medium code length. Surprisingly, we don't see a pattern in fixating on \texttt{ClassDef} in all different code lengths. For the Procedural paradigm, we observe that in both the Large and Small code lengths, most of the participants fixate on \texttt{Return, arg, FunctionDef} frequently while reading. This is not the case for \texttt{Return} in the Medium code length, which could be because there is only one return statement, and it is placed at the end of the code. 
As for the mixed paradigm, we did not notice clear patterns of what token types were fixated on more frequently, as the participants looked at different relevant and irrelevant token types for each code length.
While in \Cref{fig:classify_m_l}, the Large Mixed paradigm code, we see participants fixating mostly on relevant token types during their reading, we do not see similar behavior when looking at the Medium and Small code length plots, which show fixations on both relevant and irrelevant tokens.

Our observations indicate that when participants read code for classifying the predominant paradigm, they fixate on all types of tokens and are not specifically looking for specific token types. In the Functional paradigm, the token types that were fixated on the longest are \texttt{Name} and \texttt{Lambda}. In the Object-Oriented paradigm, \texttt{Name} was fixated on the longest, followed by \texttt{Attribute} and \texttt{ClassDef}. \texttt{Name} and \texttt{FunctionDef} were fixated on the longest in the Procedural paradigm, and finally, \texttt{Name}, \texttt{Constant}, and \texttt{FunctionDef} were fixated on the longest in the Mixed paradigm.  \texttt{Name} was the token type fixated on the longest over all the paradigms and code lengths. The token types the participants focused on seemed to depend on the code itself, not necessarily the paradigm.

\begin{finding}{}{fd:paradigm-et-fixations-on-types}
When participants read code to classify the predominant paradigm, they fixate on all token types and the tokens they fixate on appear to vary.% based on the task.
\end{finding}

\subsubsection{Classification Linearity Metrics}
\label{sec:rq2-class-linearity}

Next, we look into the gaze-based linearity metrics, which measure the level of linearity in reading source code. We explore the effects of paradigm and code length on the participants' reading patterns. We ran ANOVAs on the following metrics: vertical reading metrics (\textit{vertical next text}, \textit{vertical later text}), \textit{horizontal later text}, \textit{regression rate}, and \textit{saccade length}. The detailed ANOVA tables are all included in the replication package~\cite{replication-package}.

The ANOVA results on vertical reading metrics indicate that the code length and the interaction between paradigm and code length significantly affect the vertical code reading patterns. Further observations into the \textit{vertical next text} metric over the code length show that gazes are less likely to move forward one line in the Large code length (Large \textless{} Medium \textless{} Small) and results from the \textit{vertical later text} metric show that gazes are less likely to move forward multiple lines in the Medium code length and more likely in the Small code length. We observe that both vertical linearity reading metrics are significantly affected by the size of the task code, but not as much by the paradigm.

We look into the horizontal reading patterns by running ANOVA for \textit{horizontal later text} and find that paradigm, code length, and the interaction between paradigm and code length significantly affect the horizontal reading patterns. The results show that gazes are less likely to move forward within a line in the Procedural paradigm but more likely to do so in Small code lengths. We could not find significant differences between the groups looking at the \textit{regression rate} metric, which indicates that there are not significant differences between how gazes move backward horizontally on a line in the different paradigms and code lengths studied.

\begin{finding}{}{fd:paradigm-et-linearity}
The linearity metrics show that code length (rather than paradigm) is important in how developers read code. The vertical linearity metrics significantly differ for different code lengths, indicating different vertical movement searching patterns during tasks. 
\end{finding}

\subsection{\ref*{rq3} Results: Accuracy and Efficiency of Debugging Logical Errors}
\label{sec:rq3-results}

Next, we examine \ref*{rq3}, asking if the predominant paradigm has an effect on the ability to debug logical errors.
For this, we consider two metrics: accuracy and time-to-completion.

\subsubsection{Debugging Accuracy}
\label{sec:debugging-accuracy}

We show the judgment accuracy broken down by both paradigm and program in \Cref{tab:debugging-accuracy}.
We find that the paradigm of the code does not have a statistically significant effect on participants' ability to determine which line contains the bug (\statistic{Q(3)}{7.73684210526316}, \pvalue{0.0517752933734987}, \statistic{\eta^{2}_{Q}}{0.088929}).
However, as \pvalue{0.0517752933734987}, we perform pair-wise comparison anyway.
Bonferonni-adjusted \(p\)-values show no difference between paradigms, though there may exist a difference (\chisquare{1}[4.9], \pvalue[adj]{0.1611402}, \pvalue{0.0268567}) between Procedural (93\% accurate) and Mixed (66\% accurate).
That is, the paradigm does not influence the ability to debug code.

However, the program does appear to influence correctness (\statistic{Q(3)}{11.947348}, \pvalue{0.0075656}, \statistic{\eta^{2}_{Q}}{0.13726}),
though after pair-wise comparison we only suspect a difference between Factorial and Cube (\chisquare{1}[6.75] \pvalue[adj]{0.05624861}) and between Factorial and Palindrome (\chisquare{1}[4.08333], \pvalue[adj]{0.259}, \pvalue{0.043308143}).
It appears no matter the paradigm, participants may have found the Factorial code more difficult to debug (59\% accurate).

\begin{table}[htbp]
  \centering
  \caption{Number of correct responses to number of participants who viewed a given stimulus in bug localization. Highlighted cells represent stimuli categories that are different from the whole.}
  \label{tab:debugging-accuracy}
  \vspace{-0.5em}
  \small
  \renewcommand{\tabcolsep}{5pt}
\begin{tabular}{lrrrrr}
\toprule
 & \textbf{Cube} & \textbf{Factorial} & \textbf{Largest Number} & \textbf{Palindrome} & \textbf{Overall} \\
\cmidrule{2-5}
\textbf{Functional} & 7 / 7 & 4 / 8 & 7 / 8 & 6 / 6 & 82.76\% \\
\textbf{Mixed} & 5 / 6 & 4 / 7 & 4 / 8 & 6 / 8 & \highlight{65.52\%} \\
\textbf{Object-Oriented} & 7 / 8 & 3 / 6 & 5 / 7 & 6 / 8 & 72.41\% \\
\textbf{Procedural} & 8 / 8 & 6 / 8 & 6 / 6 & 7 / 7 & \highlight{93.10\%} \\
\midrule
\textbf{Overall} & 93.10\% & \highlight{58.62\%} & 75.86\% & 86.21\% &  \\
\bottomrule
\end{tabular}
\end{table}

Finally, we examine the relationships between confidence in the answer and presented paradigm, between confidence and program, and confidence and correctness.
Participants appeared less confident than when classifying code (64\% at least ``Confident'', compared to 88\%).
There appears to be a statistical relationship between confidence and whether a participant could select the correct line (\pvalue{1.479e-5}).
We also see a relationship between the program and confidence (\pvalue{0.002499}), suggesting that participants find some programs more difficult.
We do not see a relationship between the presented paradigm and confidence (\pvalue{0.1969}).

\begin{finding}{}{fd:debug-accuracy}
The code's paradigm may influence how accurately developers can debug, with accuracy higher in Functional and Procedural code.
Additionally, no difference in confidence between the different paradigms was found, though participants were more confident in some programs than others.
\end{finding}

\subsubsection{Debugging Efficiency}
\label{sec:debugging-efficiency}

\begin{figure}
  \centering
  \includegraphics[width=0.8\linewidth,page=2]{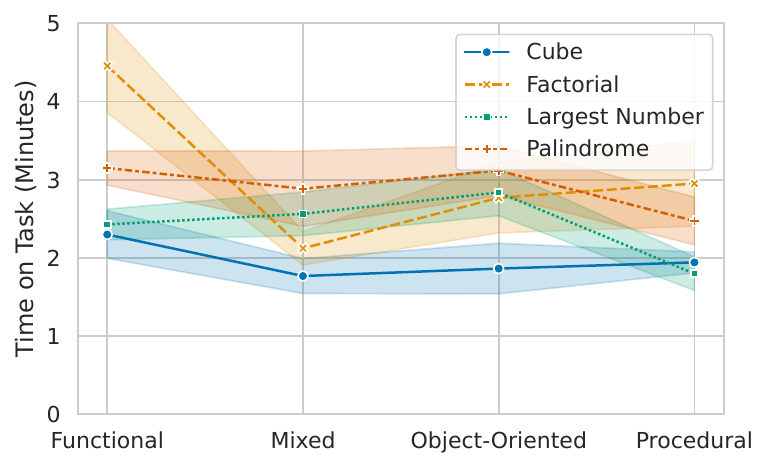}
  \caption{Average time on task, per paradigm, for debugging, including program.}
  \label{fig:debug-time-on-task}
\end{figure}

In \Cref{fig:debug-time-on-task} (above), we again present the results of a factorial ANOVA (\(4 \times 4\)) to explore the effects of paradigm and program on time to debug.
In this case, we see significant effects for the presented paradigm (\Fmeasure{3}{100}[3.995], \pvalue{0.00986}, \statistic{\eta^{2}_{p}}{0.11}, \statistic{f}{0.35}), program (\Fmeasure{3}{100}[8.45], \pvalue{4.66e-5}, \statistic{\eta^{2}_{p}}{0.16}, \statistic{f}{0.50}), and the interaction of the two.
We again perform pairwise comparisons.

We see a significant difference between Functional and Mixed paradigm code (\pvalue{0.0257}), and between Functional and Procedural code (\pvalue{0.0128}); in both cases, we see developers took longer to find the bug in Functional code.
We also see a difference between Factorial and Largest Number (\pvalue{0.0309}), as well as Factorial and Cube (\pvalue{8.84e-5}), and Palindrome and Cube (\pvalue{0.0015}).

When we consider the interaction of paradigm and program, we notice that Functional+Factorial takes the longest and is different even from the Mixed+Factorial programs (complete results of Tukey's HSD are available in the replication package~\cite{replication-package}).
This interaction follows the patterns we see for paradigm and program and may drive both effects.

\begin{finding}{}{fd:debug-efficiency}
    Functional code takes significantly longer to debug than Mixed and Procedural.
\end{finding}

\subsection{\ref*{rq4} Results: Language Feature(s) Gazed Upon When Debugging Logical Errors}
\label{sec:rq4-results}

\subsubsection{Debugging Fixation Metrics}
\label{sec:rq4-debug-fixation}

To determine which specific Python language feature(s) developers look at when debugging, we look at the influence of the paradigm and program on finding the bug by running a factorial \(4 \times 4\) ANOVA for fixation count (shown in \Cref{tab:anova-localization-fix-count}), fixation duration, and normalized mean fixation duration metrics (summarized in \Cref{tab:fixation-metrics-localization}). 
For fixation count, the ANOVA results show a significant effect for paradigm (\Fmeasure{3}{100}[7.785], \pvalue{0.0001}, \statistic{\eta^{2}_{p}}{0.19}, \statistic{f}{0.48}), program (\Fmeasure{3}{100}[13.546], \pvalue{0.0000}, \statistic{\eta^{2}_{p}}{0.29}, \statistic{f}{0.64}), and the interactions of the two (\Fmeasure{9}{100}[4.224], \pvalue{0.0001}, \statistic{\eta^{2}_{p}}{0.28}, \statistic{f}{0.62}).
Post-hoc comparison (detailed in the replication package~\cite{replication-package}) shows that Procedural and Functional paradigms are the driving factors behind the significant differences in fixation count. The tasks in the Functional paradigm had the highest fixation count, followed by Object-Oriented, Mixed, and Procedural. Cube is the program that affects the fixation count the most among other programs, with the fixation count for the program being the lowest among all paradigms. This could have also been affected by the small length of the Cube program.

Similar to our findings for code classification, we found that while our participants attended to paradigm tokens during bug localization, it was less frequently than non-paradigm tokens.
The results of a \(t\)-test and mean fixation counts on paradigm/non-paradigm tokens are shown in \Cref{tab:fixation-counts-debug}.

\begin{table}[htbp]
  \centering
  \caption{Fixation counts on paradigm and non-paradigm tokens for bug localization.}
  \label{tab:fixation-counts-debug}
  \renewcommand{\tabcolsep}{0.25em}
\begin{tabular}{lrrrrll}
  \toprule
\textbf{Paradigm} & \textbf{Paradigm Tokens} & \textbf{Non-Paradigm Tokens} & \textbf{t} & \textbf{df} & \textbf{p} & \textbf{p (adj)} \\ 
  \midrule
Functional & 22.57 & 116.41 & -5.31 & 29.57 & $<$ .001 & $<$ .001 \\ 
  Mixed & 22.73 & 73.52 & -6.19 & 42.34 & $<$ .001 & $<$ .001 \\ 
  Object-Oriented & 42.45 & 75.48 & -3.62 & 51.38 & $<$ .001 & $<$ .001 \\ 
  Procedural & 18.86 & 54.21 & -6.35 & 34.53 & $<$ .001 & $<$ .001 \\ 
   \bottomrule
\end{tabular}
\end{table}

\begin{table}[htbp]
  \centering
  \caption{High-level summary of fixation metrics for bug localization tasks (ms is milliseconds).}
  \label{tab:fixation-metrics-localization}
\begin{tabular}{lrr}
\toprule
 & \textbf{mean} & \textbf{std} \\
\textbf{Total Duration (s)} & 156.61 & 67.29 \\
\textbf{Total Duration (m)} & 2.62 & 1.12 \\
\textbf{Fixation Count} & 105.78 & 72.51 \\
\textbf{Fixation Duration (ms)} & 104,978.53 & 55,167.70 \\
\textbf{Mean Fixation Duration (ms)} & 1,244.51 & 1,163.48 \\
\textbf{Normalized Fixation Duration (ms)} & 37,299.86 & 22,706.22 \\
\textbf{Normalized Mean Fixation Duration (ms)} & 493.75 & 853.02 \\
\textbf{Total Fixations} & 105.78 & 72.51 \\
\textbf{Vertical Next Text (\%)} & 0.56 & 0.13 \\
\textbf{Vertical Later Text (\%)} & 0.72 & 0.08 \\
\textbf{Horizontal Later Text (\%)} & 0.21 & 0.07 \\
\textbf{Regression Rate (\%)} & 0.47 & 0.03 \\
\textbf{Saccade Length (px)} & 117.99 & 39.95 \\
\bottomrule
\end{tabular}
\end{table}

\begin{table}[htbp]
  \centering
  \caption{ANOVA Results for Fixation Counts on Localization Tasks.}
  \label{tab:anova-localization-fix-count}
\begin{tabular}{lrrrrr}
  \toprule
 & \textbf{Df} & \textbf{Sum Sq} & \textbf{Mean Sq} & \textbf{F value} & \textbf{Pr($>$F)} \\ 
  \midrule
\textbf{paradigm        } & 3 & 69901.00 & 23300.33 & 7.78 & 0.0001 \\ 
  \textbf{variant         } & 3 & 121635.06 & 40545.02 & 13.55 & 0.0000 \\ 
  \textbf{paradigm:variant} & 9 & 113780.19 & 12642.24 & 4.22 & 0.0001 \\ 
  \textbf{Residuals       } & 100 & 299309.93 & 2993.10 &  &  \\ 
   \bottomrule
\end{tabular}
\end{table}

Next, we investigate the effects of paradigm and program on fixation duration. Once again, we see significant differences in fixation duration between different paradigms (\Fmeasure{3}{100}[3.017], \pvalue{0.0334}, \statistic{\eta^{2}_{p}}{0.08}, \statistic{f}{0.30}), programs (\Fmeasure{3}{100}[8.490], \pvalue{0.0000}, \statistic{\eta^{2}_{p}}{0.20}, \statistic{f}{0.50}), and the interactions of the two (\Fmeasure{9}{100}[2.131], \pvalue{0.0335}, \statistic{\eta^{2}_{p}}{0.16}, \statistic{f}{0.44}). Post-hoc comparison (detailed in the replication package~\cite{replication-package}) showed significant differences between Functional and Procedural programs, with the Functional paradigm having the highest fixation duration among all paradigms, similar to the pattern of fixation count. Once again, similar to the fixation count metric, the Cube program drives the differences between programs.

The final fixation-based metric we look at is normalized mean fixation duration. We see a significant effect from the programs (\Fmeasure{3}{100}[4.568], \pvalue{0.0048}, \statistic{\eta^{2}_{p}}{0.12}, \statistic{f}{0.37}) and the interaction between programs and paradigm (\Fmeasure{9}{100}[2.001], \pvalue{0.0467}, \statistic{\eta^{2}_{p}}{0.15}, \statistic{f}{0.42}). A post-hoc comparison shows a similar pattern to fixation count and duration, with Cube affecting the metric significantly. On average, normalized mean fixation duration was higher in the Cube program, specifically in the Functional paradigm, in which this metric was the highest among all tasks and programs. The results indicate that although the Cube program is small with the lowest fixation count and duration among all paradigms, it still required relatively longer focus and more cognitive load from the participants. Additionally, even though the Factorial program had the highest fixation count and duration, its normalized mean fixation duration was, on average, similar to the Largest Number and Palindrome, further confirming the effect of the Cube program on the differences.

\begin{finding}{}{fd:debug-et-fixations-influence}
Fixation-based metrics for the debugging tasks indicate that paradigm and program significantly influence the metrics.
The Functional code has the largest effect on metrics, as does the Cube program; notably the Functional Cube task required the longest focus from the participants
\end{finding}

\begin{figure}[ht]
  \centering
  \includegraphics[width=\textwidth]{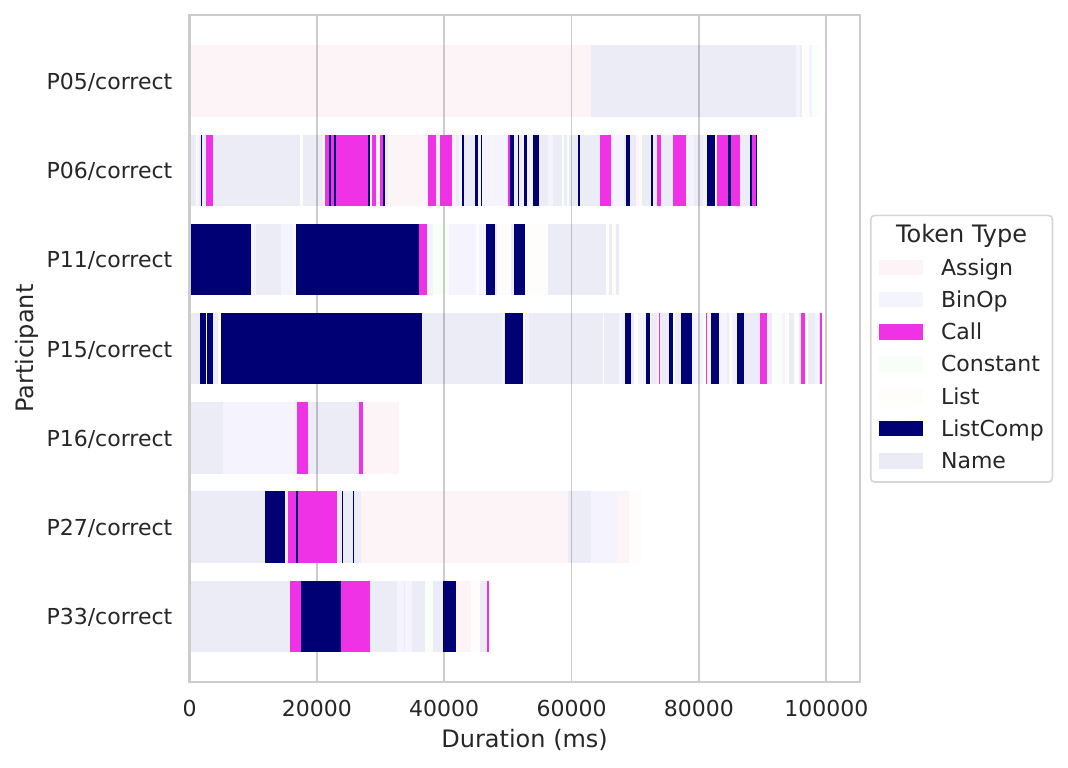}
  \caption{Scarf plots showing fixations over tokens for bug localization task, with participants labeled on \(y\)-axis (participant number and correctness) and time (in \unit{ms}) on the \(x\)-axis: Functional paradigm, Cube program}
  \label{fig:debug_c_f}
\end{figure}

\begin{figure}[ht]
  \centering
  \includegraphics[width=\textwidth]{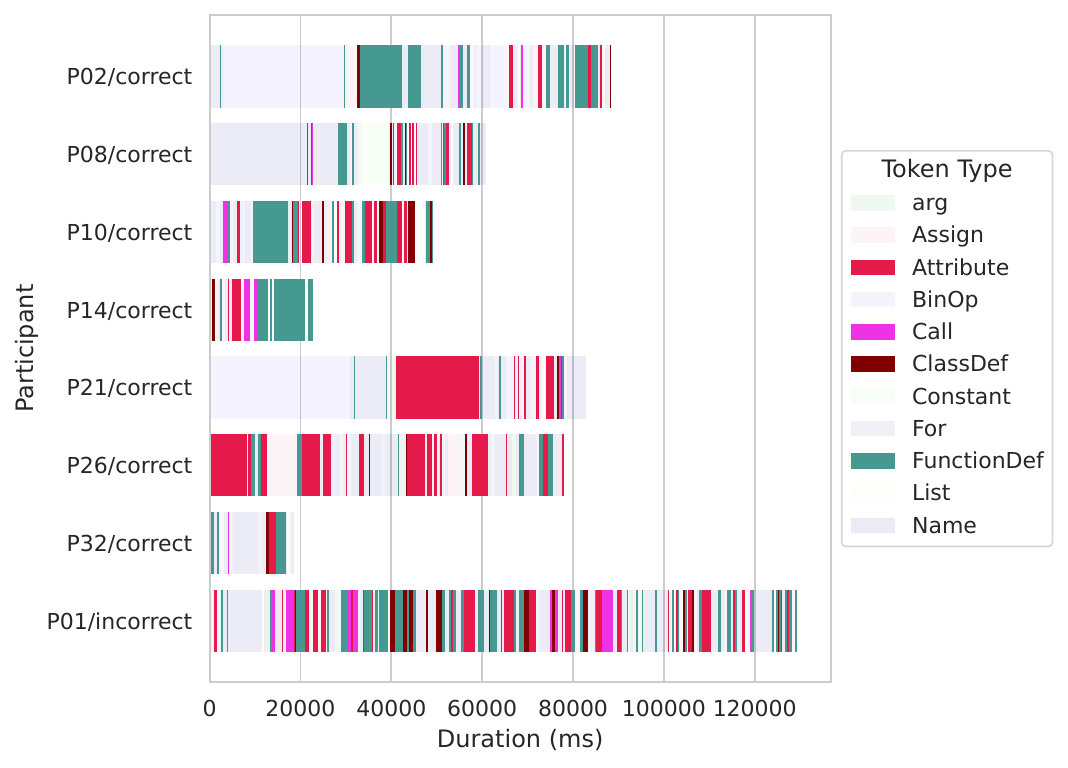}
  \caption{Scarf plots showing fixations over tokens for bug localization task, with participants labeled on \(y\)-axis (participant number and correctness) and time (in \unit{ms}) on the \(x\)-axis: Object-Oriented paradigm, Cube program}
  \label{fig:debug_c_oo}
\end{figure}

\begin{figure}[ht]
  \centering
  \includegraphics[width=\textwidth]{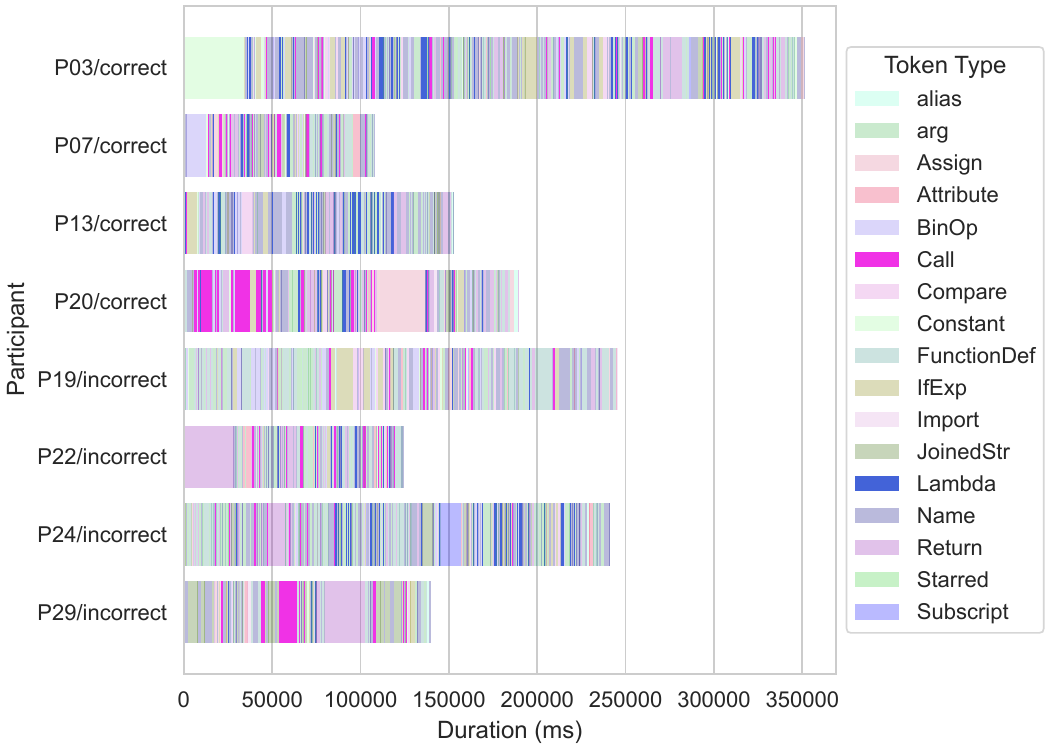}
  \caption{Scarf plots showing fixations over tokens for bug localization task, with participants labeled on \(y\)-axis (participant number and correctness) and time (in \unit{ms}) on the \(x\)-axis: Functional paradigm, Factorial program}
  \label{fig:debug_f_f}
\end{figure}

\begin{figure}[ht]
  \centering
  \includegraphics[width=\textwidth]{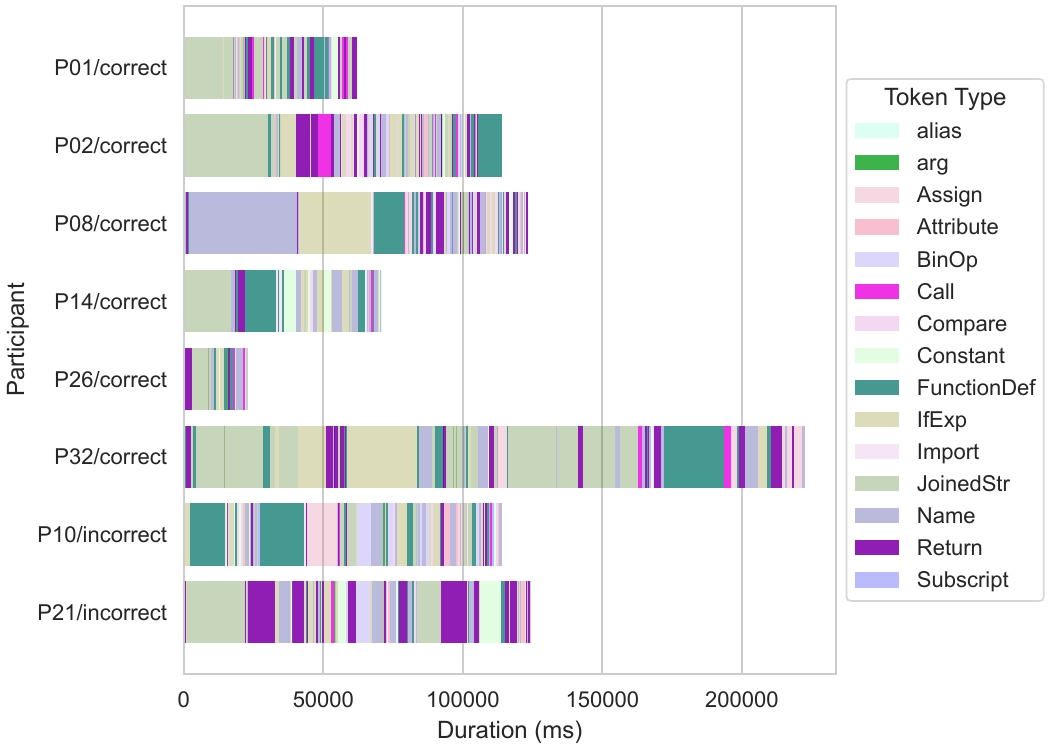}
  \caption{Scarf plots showing fixations over tokens for bug localization task, with participants labeled on \(y\)-axis (participant number and correctness) and time (in \unit{ms}) on the \(x\)-axis: Procedural paradigm, Factorial program}
  \label{fig:debug_f_p}
\end{figure}

\subsubsection{Visualization of Debugging Fixation Metrics}
\label{sec:rq4-viz-debug-fixation}

To visualize the differences in fixations, we once again use scarf plots showing in \Cref{fig:debug_c_f,fig:debug_c_oo,fig:debug_f_f,fig:debug_f_p} the fixation duration over different token types, the remaining twelve are available in our replication package~\mbox{\citep{replication-package}}.
Post-hoc pairwise comparison of the fixation metrics showed us significant differences between the Functional and Object-Oriented paradigms of the Cube program and the Functional and Procedural paradigms of the Factorial program. 
First, we look at how participants fixated over tokens while solving the Cube task in the Functional paradigm (\Cref{fig:debug_c_f}) versus the Object-Oriented paradigm (\Cref{fig:debug_c_oo}). We notice that while the participants in both groups have fixated on a lot of paradigm-relevant token types, the fixations on these tokens are relatively longer in the Functional paradigm. The longer fixations indicate that participants needed more time and focus to understand the Functional program of Cube and find the error.

Next, we look at the scarf plots for the Functional and Procedural paradigms of the Factorial program.
In the Functional (\Cref{fig:debug_f_f}) paradigm, while the participants fixated on the paradigm-relevant token types frequently, they did not fixate on these tokens longer than the non-relevant tokens (this can be explained by this task's wider variety of token types, compared to the Cube program). In the Procedural (\Cref{fig:debug_f_p}) paradigm, despite the wider variety of token types, participants spent more time focusing on the paradigm-relevant token types, especially \texttt{FunctionDef} and \texttt{Return}.

Finally, we look at what specific tokens were fixated on during debugging.  
Except in the case of Procedural, \texttt{Name} tokens were fixated upon the most (in Procedural, \texttt{JoinedString} was fixated on more).
In the Functional code, developers fixated upon \texttt{ListComp} fourth most often, though \texttt{FunctionDef} was fixated upon second most often.
In Object-Oriented, developers fixated upon \texttt{Attribute} second most often, with no other paradigm-specific tokens found in the top five.
Procedural code saw \texttt{FunctionDef} and \texttt{Assign} as the fourth and fifth most common, respectively.

\begin{finding}{}{fd:debug-et-fixations}
We found that both paradigm and program significantly influence fixation count and duration for debugging tasks, highlighting their importance in debugging. 
\end{finding}

\subsubsection{Debugging Linearity Metrics}
\label{sec:rq4-debug-linearity}

We looked into the gaze-based linearity metrics to understand how paradigm and program affect them during debugging. We ran ANOVAs on the vertical reading metrics (\textit{vertical next text} and \textit{vertical later text}) and found that paradigm, program, and their interaction significantly affect these reading patterns. Specifically, by looking at the post-hoc comparison, we observe that these significant differences are driven by the Functional paradigm and the Cube program, which is similar to our findings with fixation metrics. Both metrics show that gazes are most likely to move forward one or multiple lines in the Functional paradigm (Functional \(>\) Procedural \(>\) Mixed \(>\) Object-Oriented). The \textit{horizontal later text} follows the same pattern as the vertical metrics. While the \textit{saccade length} is largest in the Functional paradigm, it's smallest in the Procedural paradigm (Functional \(>\) Object-Oriented \(>\) Mixed \(>\) Procedural). Finally, the \textit{regression rate} metric is affected only by the paradigm,
with significant differences between the Functional paradigm and other paradigms, indicating that gazes are less likely to move backward in the Functional paradigm code. The detailed ANOVA tables are all included in the replication~\cite{replication-package}.

\begin{finding}{}{fd:debug-et-linearity}
  For debugging tasks, gaze-based linearity metrics, particularly vertical next text and vertical later text, are significantly influenced by paradigm and program, with the Functional Cube program showing distinct reading behavior, characterized by forward movements.
  In particular, the Functional paradigm showed many single-line and multi-line forward movements.
\end{finding}

\section{Discussion}
\label{sec:discussion}

This study was conducted on two types of tasks on Python code --- paradigm classification and debugging within a specific code paradigm in Python. 
Our results show interesting differences in reading behavior and performance for the Functional code across both the paradigm classification and bug localization tasks.
In the next section, we first interpret our results for each research question. 
These differences in reading behavior and differences noted more generally, between Functional code and code in other paradigms, have some interesting downstream impacts, particularly on theory building, education, research, and practice.
The rest of the section discusses these impacts.

\subsection{Interpretation of Results}
\label{sec:interpr-results}

The study investigated four research questions that resulted in eleven unique findings as described in the results.
Below, we seek to distill some high level observations from those findings.
  First, we note that we always distinguish task (classification or debugging) and stimuli (the specific program given to developers).
  This said, we offer the following explanations for our results.

    When we consider our findings for RQ1 and RQ2 (\Cref{fd:confusion,fd:paradigm,fd:classification-slower,fd:paradigm-et-fixations,fd:paradigm-et-fixations-on-types,fd:paradigm-et-linearity}), we see a potential explanation.
    The task of code classification does not require a particularly in-depth understanding of code; instead, the driving factor is to locate paradigm-specific tokens, even if they are not attended as frequently as non-paradigm tokens (\Cref{fd:paradigm-et-fixations-on-types})
    The drive to detect paradigm-specific tokens may also explain why certain paradigms are easier to classify, as participants may have been more familiar with certain paradigms than others (as they are more common or have more memorable keywords (\Cref{fd:confusion})) or certain paradigms might structure the code in a manner that is more familiar to participants (such as Object-Oriented code having a standard structure starting with a class declaration and then indented methods).
    This may also explain the differences in reading linearity driven by length (\Cref{fd:paradigm-et-linearity}).
    This same knowledge issue may also explain why some paradigms took longer to classify (\Cref{fd:classification-slower}), however, the generally uniform confidence of the participants (\Cref{fd:paradigm}) may be instead due to feeling comfortable with the classification schema.

During debugging, it is unsurprising that paradigm influences reading linearity (\Cref{fd:debug-et-fixations-influence,fd:debug-et-linearity,fd:debug-et-fixations}): paradigm influences how code is written, particularly in a language like Python, where definition order is important.
Because localizing a bug requires a deeper understanding, it is generally necessary to follow a program's logic, often following the data or control flow, which may more or less be top-down in different paradigms.
This is especially notable for Functional code, which exhibited a large amount of forward movement, following very linear code.
The higher fixation counts, longer fixation durations, and high vertical and horizontal reading patterns are indicative of high cognitive load.

Our results also indicate that procedural code had the highest accuracy for finding bugs (\Cref{fd:debug-accuracy}).
This is indicative of the fact that even though participants knew Python, the functional paradigm in Python was not as familiar to them as the other paradigms, and may also have impacted the code classification tasks.
These results should indicate that we need to consider changing how we teach Python.
As software engineers, we are often told to use the right tool for the job.
Perhaps in some cases, the Functional paradigm usage in Python is the most efficient way to solve a problem, but if it carries with it a high cognitive load (for whatever reason --- perhaps lack of knowledge structures for comprehension), it makes it harder for someone without appropriate training to comprehend it.

\subsection{Impacts on Education}
\label{sec:impacts-education}

The observed differences in code classification correctness are especially important to the education field.
We saw that participants clearly understood one paradigm (Object-Oriented code), but appeared confused between Functional and Procedural code.
This is especially notable as, in Python, only 40\% of files were previously found to be predominantly Object-Oriented~\citep{Dyer_2022}.
This in particular could be caused by a lack of familiarity with these paradigms in general or in how they are exhibited in Python.
In either case, explaining the different paradigms and showing their use to students early may help to reduce differences in classification correctness, and over time, may induce changes in how developers classify and read code.

\subsection{Impacts on Research}
\label{sec:impacts-research}

When we consider bug localization, the differences in accuracy between paradigms are lessened, though mixed-paradigm code has the lowest percentage of correct answers.
This does not follow with the differences in linearity. Thus, understanding the connection between the reading linearity and the act of debugging will require further research, especially given that some paradigms, such as Functional, involve more and larger movement as well as being uncommon in code.
In particular, determining if features of Functional code are atoms of confusion may be particularly helpful~\citep{dacost23}.

\subsection{Impacts on Practice}
\label{sec:impacts-practice}

% Practice 

Finally, because we found that developers tend to perform poorly in Mixed-paradigm code, we believe that developers should be careful in how they mix different paradigms.
In particular, it is important that developers make sure the paradigms are well-matched. What that looks like for a given project should be specified in its style guide.
  However, given our primarily-student population, further research is necessary to determine the effect of mixing paradigms on professional developers.
  Additional research is likewise necessary to determine what a style guide could look like for mixed paradigm code.

\subsection{Impacts Towards Future Theory}
\label{sec:impacts-theory}

The program comprehension literature reviewed by Storey et al. almost two decades ago~\citep{SQJ-Storey06, iwpc-Storey05} presents the main program comprehension theories as presented early in the 1960's when the procedural paradigm was prevalent. These theories include the use of beacons and chunking~\citep{Beacons}, top-down vs. bottom-up comprehension~\citep{LETOVSKY1987325, BROOKS1983543}, and programming plans~\citep{iwpc-Storey05}. These theories suggest that developers use prior knowledge bases and structures to understand programs. In 2017, Siegmund et al. use neuroscience methods (fMRI, in this case) to validate some of these theories and showed beacons ease comprehension but did not find support for programming plans~\citep{NeuralProgramComprehension}. The reasons for this could be many one being the low expertise of participants. They call for more studies to validate this theory in future work.

We posit that theories such as cognitive load theory~\citep{CognitiveLoadHCI, CognitiveLoadCompEd, CognitiveLoadEyeTracking} could also be directly related to why the functional paradigm took longer in our study. 
These theories need to be studied in relation to program comprehension of mixed paradigm code. Gaze behavior in itself might not be the only way towards building an underlying theory about how developers work with Python code written in mixed paradigms. But it can certainly supplement other methods (surveys, think-aloud, interviewing) of gathering evidence. Gaze gives us fine-grained details of where developers look while solving a task and shows us the thought processes being used to develop their mental model. We view this study as one of the first steps in understanding different paradigm usage in Python.

With the addition of newer paradigms and the use of coding assistants like ChatGPT and Copilot, a fresh look into whether prior theories apply to program comprehension tasks is needed. Program comprehension research should look into whether these existing theories are still valid or need refinement as part of this culture shift. As we gather more evidence in this area of mixed paradigm languages, we help gather evidence towards building a theory of mixed paradigm comprehension and debugging. These theories can also help program language designers as they work on new languages. For example, what can language designers incorporate to make functional languages more user friendly? We are not aware of any theories claimed to be used in programming language design.

We can anticipate a theory that functional programming takes longer since it requires longer time to build the mental structures indicating higher cognitive load~\citep{CognitiveLoadCompEd}, to understand what the program does. There are many other factors that could also play in role in the way developers comprehend different paradigms related to knowledge structures, prior work experience, and other socio-technical factors. As more eye tracking studies are conducted with Python, researchers will be able to extract common phenomena observed in developer behavior that can eventually be used to formulate a theory of program comprehension for mixed paradigm languages such as Python.  Such a theory would involve many factors including the developers behavior, processes used, paradigms, experience, prior knowledge,  project domain and more.

\section{Threats to Validity}
\label{sec:threats}

We discuss potential threats to the validity of our study and how we mitigate them.

\paragraph{Construct Validity.}
First is the threat of start/stop time measurement.
These measurements are derived from when the participant first starts looking at the target file and when they stop reading it. There may be a small lag in stop times due to the manual nature of stopping recording upon participant completion. As this lag is fairly consistent for all tasks, we do not believe it affects the results.
How the paradigm is defined bears some influence on responses as participants may have a different internal definition of a paradigm and the definition used may be confusing.
This is especially notable as participants' prior experience potentially informs their internal definitions so that prior experience may have biased their answers. Our background survey failed to collect data on participant experience with each paradigm, and results may be biased.
Moreover, as different syntactic features may co-occur, this may cause other problems. We remedy this by consistently using definitions from prior literature~\cite{Dyer_2022}.
Additionally, we used a tutorial to train participants in the definitions of the language features used.
The code presented is fairly linear, which may not accurately reflect how all paradigms are used in the real world.

We use fixation counts and fixation durations for our eye tracking metrics for RQ2 and RQ4.  An alternate method could use the fixations per second (fixation rate) metric for fixation counts. The number of fixation counts are inherently dependent on the time on task, especially since the time is different for each participant (per task) as in our study. We mitigate this risk by using fixation durations (also related to task time) and normalizing them by dividing by the length of the fixated-on token.

Finally, the study was set up to provide good calibration on the presented code (accessible font size) using a high-quality research-grade tracker. All the tasks were less than 5 mins in length with recalibrations done between tasks. Each task was also recorded separately. In addition, the moderator made sure the tracking was done accurately by watching a live view of the participants' head position and gently prompting them to adjust when needed. Because of the above stated reasons, we did not find the need to correct fixations. Automatically correcting fixations does require a validated golden set as shown by~\cite{etra25-srcGaze} for Java source code.

\paragraph{Internal Validity.}
There are several threats to internal validity.
The first threat is our fixed order of presentation for classification.
Classification tasks were always presented as ``Functional'', ``Object-Oriented'', ``Procedural'', then ``Mixed''.
This fixed order may have introduced two possible threats: fixed order may confound the relationship to paradigm, and fixed order may have induced a learning effect.
This is a particular threat to our conclusions on correctness and time to completion on classification.
However, because of the fixed order of task, and un-fixed order of paradigm presentation in bug localization, we can confirm at least some of these differences, especially as relates to eye-tracking-derived metrics.

Finally, our explicit training on the paradigm classification presents a threat to validity.
Because we explicitly train participants, this can lead to a mismatch and bias on the measures related to correctness or time-on-task.
However, because of how the training is completed, we do not believe that the eye-tracking-specific measures are likely to be impacted.

\paragraph{External Validity.}
Our study had primarily student participants, limiting generalizability.
In particular, students' limited experience with multi-paradigm code, or any given paradigm, limits the ability to generalize to developers with broader experience.
However, as this study is primarily exploratory in nature, the use of convenience sampling is sound.

Another threat concerns the tasks themselves. As the tasks used were short and simple in terms of complexity, the ability to generalize to more complex code may be at stake. 
However, despite the simplicity, this limitation may matter less than expected as we see a number of differences.
The tasks are all written in Python, so results may not generalize to other multi-paradigm languages.

  Moreover, the small number of tasks further reduces generalizability, as relatively few paradigm-related constructs were studied.
  However, the small number of tasks increases both internal and conclusion validity, allowing us to better compare results across paradigms.

Additionally, the varying sizes (and designs) of the tasks across paradigms reduces our ability to truly generalize the results with respect to difficulty or time.
However, the subjective results of perceived difficulty should remain a strong indicator.
Additionally, although there is variance in task size, all tasks remain quite small, which improves comparability but does reduce generalizability to larger tasks.

\paragraph{Conclusion Validity.}
Our statistical analyses are well-supported by our experimental design.
First, since the variables analyzed are binary (correctness), mutually exclusive, matched (or paired in the case of the \(\chi^{2}\)) and independent per subject, the use of the non-parametric \(Q\) with McNemar's \(\chi^{2}\) to follow-up is supported.
Moreover, our use of Bonferroni adjustment reduces the risk of Type I error in this circumstance.

Next, our use of Fisher's exact test to analyze relationships between confidence and the presented paradigm, length, or program is similarly appropriate since the data analyzed is categorical (and in some cases ordered), with small, fixed marginal totals.

Additionally, our sample size (at 28 retained participants) is small.
However, for an observed effect size of \(f = 0.30\), (fixation duration on localization, difference between paradigm), we have a theoretical power of 74.7\%.
Although 74.7\% power is not ideal, other effect sizes are higher, and at \(f = 0.35\), we have 87.7\% power, which is substantially better.
That is, we generally have at least sufficient power to detect our effects.

Finally, our use of Factorial Mixed-Groups ANOVA is supported since our dependent variables are continuous and grouping variables are as required (i.e., within-groups variables are truly within-groups, and between-groups variables are truly between-groups).
We do not see any outliers in the data, and mixed-group models are known to be robust to minor deviations from normality.
As Tukey's HSD requires similar assumptions, it is likewise supported.
There may, however, be other confounding factors, such as the size of the code across paradigms.

\section{Conclusions and Future Work}
\label{sec:conclusion}

In this study, we used eye tracking to first determine if developers look at particular language features when attempting to classify the predominant paradigm in Python code, and second, if language features played a role in isolating bugs. 
We found that developers tend to take longer to classify code written in the functional paradigm, but their confidence does not change based on the shown paradigm. 
Developers also incorrectly classified procedural code, likely because they confused procedural and functional constructs.
When classifying code, we found that participants fixated on all types of tokens, not only those specific to the paradigm of code they were shown.
Moreover, we found that paradigm does not strongly influence developers' ability to isolate bugs, though participants were overall less confident.
One paradigm in particular (Functional) did cause participants to take longer to isolate bugs. However, the program they were debugging also influenced this.
Finally, fixation-derived metrics (including linearity) were affected by Functional code which was read in a more linear order than other paradigms.

In the future, we want to better understand why developers consider code as being mixed and not predominantly in one paradigm. Additionally, we want to explore the connection between atoms of confusion, code smells, and programming paradigms.

\section*{Data Availability}
All de-identified data and processing scripts are available in our replication package~\cite{replication-package} on Zenodo.

\section*{Compliance with Ethical Standards}

\subsection*{Conflict of Interest}

The authors declare that they have no conflict of interest.

\subsection*{Human Participants and Informed Consent}

This experiment was reviewed and approved by the University of Nebraska--Lincoln Institutional Review Board, as project \#21455, IRB number \#20220121455EX.
Study participants gave informed consent and were able to withdraw consent at any time.

\subsection*{Funding}

This material is based upon work supported by the U.S. National Science Foundation under grant numbers 23-46327 and CCF 18-55756.

\begin{acknowledgements}
  We thank all the participants of this study for their time. 
\end{acknowledgements}

\clearpage
\bibliographystyle{spbasic}
\bibliography{references}

\end{document}